\documentclass{article}

\usepackage[utf8]{inputenc}
\usepackage[margin=1in]{geometry}
\usepackage{hyperref}
\usepackage{natbib}
\usepackage[USenglish]{babel}
\usepackage{csquotes}
\usepackage{lscape}
\usepackage{mathtools} 
\usepackage{amssymb}
\usepackage{bm}
\usepackage{graphicx}
\usepackage{booktabs}
\usepackage[dvipsnames]{xcolor}

\newcommand{\RKt}{RK_{t+1d}^{\br{d}}}

\newcommand{\RKy}{RK_{t}^{\br{d}}}
\newcommand{\RKw}{RK_{t}^{\br{w}}}
\newcommand{\RKm}{RK_{t}^{\br{m}}}

\newcommand{\RKCondFour}{\RKt,\, \RKy, \, \RKw, \, \RKm}
\newcommand{\RKCondThree}{\RKy, \, \RKw, \, \RKm}

\DeclarePairedDelimiter\autobracket{(}{)}
\newcommand{\br}[1]{\autobracket*{#1}}

\DeclareMathOperator*{\plim}{plim}
\DeclareMathOperator*{\E}{\mathbb{E}}
\DeclareMathOperator*{\Prob}{P}

\newcommand\B{\rule[-1.4ex]{0pt}{0pt}}

\title{A Vine-copula extension for the HAR model}

\author{Martin Magris}

\date{Tampere University\footnote{Faculty of Information Technology and Communication Sciences (ITC), Tampere University, P.O. Box 541, FI-33101 Tampere, Finland. Email: martin.magris@tuni.fi.} \footnote{A significant part of this research and manuscript were elaborated during author's visiting period at the CREATES center, Aarhus University, Denmark (2018). An earlier version of this paper circulated in November 2018.}\\[2ex] July, 2019}
\begin{document}

\maketitle

\begin{abstract}
The heterogeneous autoregressive (HAR) model is revised by modeling the joint distribution of the four partial-volatility terms therein involved. Namely, today's, yesterday's, last week's and last month's volatility components. The joint distribution relies on a (C-) Vine copula construction, allowing to conveniently extract volatility forecasts based on the conditional expectation of today's volatility given its past terms. The proposed empirical application involves more than seven years of high-frequency transaction prices for ten stocks and evaluates the in-sample, out-of-sample and one-step-ahead forecast performance of our model for daily realized-kernel measures. The model proposed in this paper is shown to outperform the HAR counterpart under different models for marginal distributions, copula construction methods, and forecasting settings.
\end{abstract}

\section{Introduction}
Volatility estimation and forecasting have been a major research area in financial econometrics. In the last decades, the availability of high-frequency financial data led to prolific research on volatility estimation in high-frequency settings, in particular to the development of the so-called realized measures \citep[see][for a review]{mcaleer2008realized}. 
From \citep{merton1980estimating} who first suggested that the volatility can be estimated arbitrary well employing finely sampled high-frequency returns, \citep[][among the first ones]{andersen1998answering,andersen2001distribution,meddahi2002theoretical} developed the theory of the widespread realized variance (RV) estimator for the integrated variance. The early studies of \citep[e.g.][]{andersen2003modeling,andersen2004analytical} show that indeed simple models of realized variance outperform the popular GARCH and other SV models in out-of-sample forecasting.


The slowly decreasing autocorrelation and long persistence in squared returns, along with their slow convergence to the normal distribution associated with fat tails and leptokurtic return distributions constitute well known stylized facts. These challenges for the econometric modeling empirically outline the importance of long-memory dependencies in markets' volatility \citep[see e.g.][among many others]{cont2005long}.

Several ARCH and SV models have been specifically formulated to deal with this phenomenon, usually by incorporating long-memory patterns through fractional differencing. Fractionally integrated long-memory formulations (e.g. ARFIMA or FIGARCH models) are generally complex, lacking intuitive economic interpretation and mixing long and short memory features of difficult disentangling \citep{comte1998long}. The estimation is not straightforward and requires long estimation windows \citep[e.g][]{bollerslev1996modeling}.
In forecasting high-frequency volatility measures, the heteroskedastic auto-regressive (HAR) model of \citet{corsi2009simple} stands out as the main tool, attractive for its simplicity in construction, interpretation, and estimation. Importantly, the HAR model effectively approximates the long-range dependence observed in volatility series and is able to reproduce several stylized facts. Indeed the aggregation as a sum of different processes, like the one the HAR and the earlier HARCH model of \citep{muller1997volatilities} consider, conveys long-memory features \citep[e.g.][]{granger1980long,lebaron2001stochastic}.

For its linear structure, immediate OLS estimation and remarkable performance in out-of-sample analyses, the HAR model constitutes a well-established benchmark for volatility forecasting with realized measures. Several extensions to the HAR model have been proposed. \citep{andersen2007roughing} includes a jump component in the regressors, showing that short-lived bursts in volatility are associated with jumps in the process. By following the results of \citep{barndorff2008measuring} decomposing the RV in semi-variances due to positive and negative returns, \citep{patton2015good} includes asymmetries based on the return sign, noting that negative returns are of greater impact on RV and have longer persistence than positive ones. \citep{corsi2012discrete} accounts for both the continuous and jump components of RV and leverage effects too. In a similar specification \citep{liu2007there} finds strong empirical evidence of structural breaks in realized variance. Further models and applications allowing for structural breaks include \citep{mcaleer2008multiple,hillebrand2010benefits,mcaleer2011forecasting,wen2016forecasting,gong2018structural}. Structural breaks and leverage effects are tackled under a MEM model perspective in \citep{gallo2015forecasting}.
Further extensions allow for time-varying parameters. Among them, \citep{bollerslev2016exploiting} implicitly reaches time-variation by accounting for the measurement error between the RV and the integrated variance. In \citep{chen2018nonparametric} the time variation is free of a functional form but locally approximated with a kernel function.

Recently, \citep{buccheri2017hark} introduced an autoregressive model with time-varying parameters driven by an autoregressive component and scores of the conditional density. Their time-varying specification can be employed as an alternative representation for general non-linear autoregressive specifications \citep[in this regard see][]{blasques2014optimal}. Indeed, the structurally non-linear smooth transition model of \citep{mcaleer2008multiple} with multiple volatility regimes is shown to be outperformed in out-of-sample forecasting. Considering non-linearities is an important aspect: long-memory features can be misinterpreted as non-linearities and the other way around \citep[][and references therein]{mcaleer2011forecasting}, inflating coefficient's estimates. Hence, the relevance of non-linear models embedding long-memory features for disentangling these two aspects. \citep{hillebrand2010benefits} proposes a neural network extension where a mixture of logistic functions approximates the unknown function linking log-RV and state variables, extended to dummies for weekdays, macroeconomic announcements and cumulative returns. Lagged variables are selected via bagging, a predictor selection strategy shown to improve, to different extents, the forecasting accuracy for all the models therein investigated. \citep{mcaleer2011forecasting} further expands \citep{hillebrand2010benefits} by randomly selecting the number of sums in the bagging algorithm. Similarly, but for a fixed set of predictors \citep{arneric2018neural} discusses different neural networks
alternatives for different HAR specifications, having evidence of statistically significant in-sample and out-of-sample outperformance over their respective linear counterparts.

Besides the particular vector of regressors $\bm{X}_t$ for a day $t$, inclusive or not of possible leverage or jump-variation terms, the problem of determining a suitable functional form $f$ linking $RV_t$ and $\bm{X}_t$ is challenging. This can be either retrieved by a functional approximating strategies based on machine-learning inspired methods like \citep[e.g.][]{hillebrand2010benefits,lebaron2018forecasting}, or with time-varying parameters leading to specifications of equivalent nonlinear representations for some unknown function $f$ \citep{buccheri2017hark}. 
A general regression problem takes form $RV_t = f\left(\bm{X}_t,\bm{\beta } \right)$, where $\E\left[ RV_t|\bm{X}_t\right] = f\left(\bm{X}_t,\bm{\beta}  \right)$ 
and $\bm{\beta} $ is a vector of parameters. Conversely, linear HAR-like specifications can generically be reduced to a form such as $f\left(\bm{X}_t,\bm{\beta}  \right) = \bm{X}_t^{'}\bm{\beta}$. 
In this work $\E\left[ RV_t|\bm{X}_t\right]$ is directly achieved from the conditional distribution $F_{RV_t|\bm{X}_t}$, retrieved from the joint distributions $F_{RV_t,\bm{X}_t}$. 
This formulation does not restrict the regression over a particular $f$, but allows for general functionals, implicitly determined by the joint distribution $F$, driven by the underlying copula and the complexity of the dependencies between the regressors. 
Copulas are invariant under transforms of the margins \citep[e.g.][]{nelsen2007introduction}. For alternative log- and square-root HAR specifications, the joint is readily obtained in virtue of Sklar's theorem by simply updating the margins. Importantly, conditional expectations for strictly positive multivariate distributions are by construction nonnegative: forecasts are always positive.
The rich information on the joint distribution $F_{RV_t,\bm{X}_t}$ is readily accessible when estimating any of the HAR model formulations. Yet, this information has not been so far exploited, and there are no copula-based approaches in the HAR-related literature.
Equivalence in-sample and out-of-sample forecasts wrt. the HAR model would implicitly unveil whether HAR's linear assumption is perhaps misspecified or not. 
Results favoring our specification would indicate that the information conveyed in the join distribution is by itself highly informative for tomorrow's volatility, directly exploitable without further underlying assumptions.

This paper revisits the HAR model retaining its original formulation involving three interacting volatility components at different time scales, but models their joint distribution with copulas and extracting one-step-ahead forecasts accordingly. The closest work is the bivariate framework of \citet{sokolinskiy2011forecasting}. We extend it in a setting fully resembling the spirit of the HAR model and adopt a more flexible distribution modeling approach. We adopt the recent advances in multivariate modeling provided by the so-called Vine copulas \citep[e.g.][]{joe1994multivariate,joe2011dependence}.
Setting apart from \citet{sokolinskiy2011forecasting} conditional distributions and expectations are retrieved via numerical integration over multivariate distributions, without relying on simulation. Some of our results are further discussed wrt. a simple neural network benchmark model.

The empirical application considers more than seven years of high-frequency transaction data for 10 stocks and implements different estimation and forecasting schemes, exploiting several different measures to asses the performance of our model with respect to the standard HAR. The model we develop seems to outperform the HAR model both in-sample and in one-day-ahead forecasting. 

Section \ref{sec:Volat} shortly introduces the realized measure used in our applications and recalls the concept of realized variance. 
Section \ref{sec:HAR} introduces the HAR model and motivates the non-linear copula-based model presented in Section \ref{sec:CVHAR}.
Vine copula construction is presented in Section \ref{sec:Vines}.
Section \ref{sec:EmpApp} merges all the earlier discussion, and specifies the setting of the empirical application. Section \ref{sec:Results} discusses the results, while Section \ref{sec:Concl} concludes. Figures (and tables on test statistics) are conveniently organized in the Appendix.

\section{Volatility estimation with high frequency data}\label{sec:Volat}
\subsection{Realized variance}
A major problem in high-frequency econometrics consists of the non-parametric estimation of the volatility of a price process. The advantage of rich tick-by-tick data allows in the high-frequency setting to accurately estimate the so-called integrated variance (IV). 
A widely employed specification models the logarithmic price according to the continuous time diffusion:
\begin{equation}\label{eq:defLogPriceProc}
p_{t} = p_{0} +\int_t^{t}\mu_s ds + \int_0^{t} \sigma_s dW_s
\end{equation}
Note that this specification involves a time interval $[0,t]$, say e.g. a day.
Despite the specific problem considered, a number of further operational assumption can be taken. Equation \eqref{eq:defLogPriceProc} is commonly required to be such that the variation of the drift term is neglectful with respect to the stochastic integral, $\sigma_s$ is assumed to be positive,  have a continuous sample path, and be independent of the Brownian motion $W_s$. Often $\mu_s \equiv 0$ constant, or predictable and of finite-variation. 

The object of interest in the realized variance theory is the so called \textit{integrated variance}. Let $r\left(0,t\right)$ be the compound return over the period $\left[0,t \right]$, and  $\mathcal{F}_t = \left\lbrace \mu_{s}, \sigma_{s} \right\rbrace_{s \in \left[0,t \right]}$ the $\sigma$-algebra generated by the sample paths of drift and diffusion processes. The integrated variance is defined as:
$$
IV_{t} = \int_0^t\sigma^2_{s}ds = \text{Var}\left(r \left(0,t \right) \mid \mathcal{F}_{t} \right)
$$
This is a key-ingredient commonly taken as and adequate volatility measure over the period $[0,t]$, representing a synthesis of the volatility path through the time interval under consideration.\\
Suppose the log-price process is observed over $[0,t]$. For convenience the price is sampled at regular sub-intervals of size $\delta$, be $\lbrace p_0,...,p_{i\delta},...,p_{n\delta} \rbrace$  the observed prices, ${i =  1,...,n }$ and $n\delta=t$. The \textit{realized variance} (RV) is defined as:
\begin{equation}\label{eq_defRV}
RV_{t,n}=\sum_{i=1}^n\br{p_{i\delta}-p_{\br{i-1}\delta} }^2 = \sum_{i=1}^n r^2{\left((i-1)\delta,i\delta\right)}
\end{equation}
The realized variance is in fact the second sample moment of the return process over the interval, scaled by the number of observations to provide a measure calibrated to the length of the measurement interval \citep{andersen2010parametric}.

A well-known implication is that the realized variance is a consistent estimator for the increments in quadratic variation of a process \citep[see e.g.][]{andersen2010parametric, barndorff2002estimating}. Under eq.\eqref{eq:defLogPriceProc}, this implies that the RV  consistently estimates the corresponding IV, i.e. $\plim RV_{t,n} =IV_t$. 
At high sampling frequencies, non-negligible market microstructure noise (MMS) effects
turns the estimator biased and inconsistent. Whether one would like to sample at the highest possible frequency in virtue of the above consistency result, microstructure noise constitutes a limit. 5-minute sampling is a common threshold \citep[e.g][]{andersen1998answering}, however the realized variance is inevitably affected by discretization error \citep{barndorff2003realized,barndorff2006limit}.

Market microstructure noise (MMS) is a key concept in high-frequency econometrics. MMS is an error source contaminating the ideal price process of eq.\eqref{eq:defLogPriceProc}, dominant at high frequencies. Empirical evidence shows that the ideal model in eq.\eqref{eq:defLogPriceProc} is inappropriate when prices are sampled at high frequencies: random MMS noise affecting the price should be taken into account to consistently estimate the integrated variance with no bias \citep[e.g.][among many others]{hansen2006realized}.

\subsection{Realized kernel}
Among the feasible IV estimators developed for MMS regimes, we adopt the realized kernel (RK).
Exhaustive references on the RK are \citep{barndorff2006limit,barndorff2008designing,barndorff2009realized,barndorff2011multivariate}, while the general idea of kernel-based estimator for the integrated variance in MMS setting goes back to \citep{zhou1996high,barndorff2004regular,hansen2006realized}.

We discuss the univariate kernel and of \citep{barndorff2009realized}, whose theoretical foundations comes from \citep{barndorff2008designing}. With $r_j$ represents the $j$-th high-frequency return calculated over the interval $\left[ t_{j-1},t_j\right]$ (tick-by-tick return). The realized kernel estimator takes form:
\begin{equation}\label{eq:RK}
    RK_n = \sum_{h =-H}^H k\br{\frac{h}{H+1}}\gamma_h, \;\;\;\;\;\;\;\;\;\; \gamma_h = \sum_{j=|h|+1}^n r_j r_{j-|h|}
\end{equation}
where $k$ is a kernel weighting function. In applications the Parzen kernel is the preferred choice \citep{barndorff2008designing,barndorff2009realized}. Within the RK analyzed in \citep{barndorff2008designing} the kernel estimator in eq.\eqref{eq:RK} is the so-called non-flat-top realized kernel, which is guaranteed to produce \textit{non-negative} estimates. 
Indeed the assumption on the efficient price process is more relaxed wrt. eq.\eqref{eq:defLogPriceProc} (e.g. allowing for jumps), while the estimator is consistent also under serially-dependent noise. \citep{barndorff2008designing} also develops the theory for the selection of an optimal bandwidth $H^*$ in terms of best trade-off between asymptotic bias and variance. The estimation of the quantities involved in determining $H^*$ and the overall implementation of the RK in eq.\eqref{eq:RK}, are in detail discussed in \citep{barndorff2009realized}. In our implementation, we adopt the Parzen kernel function and use one observation at each of the sample endpoints for jittering.

\section{HAR model}\label{sec:HAR}

The HAR model of \citet{corsi2009simple} stands as a generalization of earlier HARCH models \citep{muller1997volatilities}, heuristically motivated by the heterogeneous market hypothesis, which assumes the existence of different type of agents, heterogeneous over the different investment horizons they trade. \citet{corsi2009simple} shows that a simple linear model obtained by mixing three volatility components is able to reproduce the typical slow decay in volatility autocorrelation, stylized facts about returns' and volatility distributions and, although its simplicity, has been shown to be difficult to beat in terms volatility forecasting.

The HAR model assumes a three-factor stochastic volatility model for the latent volatility, identified by the daily integrated variance, captured with an appropriate  measure\footnote{The original model of \citet{corsi2009simple} adopts the RV, but applies to general intraday realized measures, e.g. the RK \citep[][among the others]{hillebrand2010benefits,gallo2015forecasting}.}. The model assumes that the daily volatility process is a function of the past daily realized volatility and of longer-term partial volatility components \textendash daily component ($d$), weekly component ($w$), and monthly component ($m$). \citet{corsi2009simple} suggests the following  simple time-series representation:
\begin{equation} \label{eq:HAR}
    RK_{t+1d}^{\br{d}}= c + \beta^{\br{d}}RK_t^{\br{d}}+\beta^{\br{w}}RK_t^{\br{w}}+\beta^{\br{m}}RK_t^{\br{m}}+\omega_{t+1d}
\end{equation}
In particular, $RK^{\br{w}}_t=\frac{1}{5}\sum_{i=0}^4 RK^{\br{d}}_{t-i}$ and $RK^{\br{m}}_t=\frac{1}{22}\sum_{i=0}^{21} RK^{\br{d}}_{t-i}$ are respectively interpretable as weekly and monthly partial volatility terms\footnote{In eq.\eqref{eq:HAR} $t+1d$ reads as \enquote{(end of) day $t$ plus one day}.}. Volatility innovations are serially independent and zero-mean, with a truncated left tail to guarantee positivity.
Eq.\eqref{eq:HAR} corresponds to an autoregressive model with autoregressive weights taking a step-function form, restricted in a parsimonious way such that the three emerging components are economically meaningful and interpretable.

Although the HAR model stands out as the preferred choice for daily volatility modeling with intraday data, several authors have proposed modifications or improvements, e.g. by different or additional regressors \citep[see e.g.][]{andersen2007roughing,patton2015good,bollerslev2016exploiting}, or considering non-linear specifications \citep{hillebrand2010benefits,arneric2018neural}. In the following, we provide some arguments that motivate the Vine-copula research direction developed in this paper. 

The crucial feature of the HAR model is its linearity. 
Although linearity is attractive in terms of interpretability and eventually in model estimation, this stems as an assumption on the functional linkage between the components. Forecasts from the HAR model are conditional expectations day-t volatility given the observed past terms. Such conditional expectation is assumed as being a linear combination of lagged RK terms.
Moving aside from this specification and deal with potential non-linearities, instead of specifying alternative functions to link the RK terms, or applying machine learning -like algorithms to flexibly approximate an unknown functional, it appears interesting to directly look at the joint distribution between the three RK components, by focusing e.g. on their multivariate joint distribution $F_{RK_t^{\br{d}},RK_t^{\br{w}},RK_t^{\br{m}}}$, and consequently by considering the expectation of the conditional distribution $F_{RK_{t+1d}^{\br{d}}|RK_t^{\br{d}}, RK_t^{\br{w}}}$ for forecasting $RK_t^{\br{m}}$ \citep[as][suggested in a much simpler setting]{sokolinskiy2011forecasting}, without further assumptions, perhaps on the functional forms and errors' distribution of any potential model. Implicitly this framework accounts for possible non-linearities in the conditional expectation, since not constrained to a specific functional form, but directly recovered from $F$ and driven by the dependence relationships therein involved between its variables.

On the other hand, an OLS-estimated linear model like eq.\eqref{eq:HAR} is not guaranteed to produce positive estimates of tomorrow's volatility. As a turnaround, a log-specification of the HAR model can be adopted, however, the forecasts are not of direct use (Jensen inequality) and need to be e.g. bootstrapped. Also in our data, we have spurious evidence of negative estimates and confidence intervals, which are economically of no sense. In a non-log framework RK, the positiveness of RK implies a non-normal error term in \eqref{eq:HAR}. This is not affecting the OLS estimator in terms best linear unbiased estimator (BLUE) of the regression coefficients, but poses issues in inference and e.g. in predicting confidence intervals.
Non-linearity and positivity issues in the original HAR model, favor the discussion over a structural-free model that directly exploits the joint distribution of the four volatility components. Such an alternative is discussed in the next Section.

\section{CV-HAR model}\label{sec:CVHAR}
Motivated by the discussion in Section \ref{sec:HAR}, we look at the joint distribution of the RK-based day-$\br{t+1d}$, day-$t$, last week's and last month's volatility measures, $F_{\RKCondFour}$ to extract the conditional distribution $F_{\RKt | \RKCondThree}$ and compare the HAR model against the alternative:
\begin{equation}\label{eq:CVHarModel}
    \RKt = \E\left[ \RKt \vert \RKCondThree \right] + \omega_t
\end{equation}
The error term accounts for both measurement errors and the variability in $\RKt$.
Whereas the day-ahead volatility forecasts of the HAR model are obtained by evaluating the right side of eq.\eqref{eq:HAR} by plugging the observed RK values, here the forecast is the expectation of day-$\br{t+1d}$ volatility given the other volatility components begin equal to the empirically observed RK counterparts. 
To obtain the forecasts, the model requires nothing but evaluating the integral involved in the conditional expectation.
At time $t$ as a forecast for the volatility at $t+1d$, with the HAR and the above model we respectively have:
\begin{align} \label{eq:Forecasts}
    \hat{x}_{t+1d}^{\br{d}} &=  c + \beta^{\br{d}} x_t^{\br{d}} + \nonumber \beta^{\br{w}}x_t^{\br{w}} + \beta^{\br{m}}x_t^{\br{m}}\\ 
    \hat{x}_{t+1d}^{\br{d}} &=  \E \left[ \RKt \vert  \RKy = x_t^{\br{d}}, \, \RKw = x_t^{\br{w}}, \, \RKm = x_t^{\br{m}} \right]
\end{align}
where $x$ are the sampled (observed) RK values at day $t$ for the different volatility components. 

Indeed is eq.\eqref{eq:Forecasts} we should look at (rather than eq.\eqref{eq:CVHarModel}) to get the intuition behind our model: \enquote{forecast tomorrow's volatility with the conditional expectation of tomorrow's volatility given today's, last week's and last month's}. 

The model in eq.\eqref{eq:CVHarModel} can be seen a generalization of the framework in eq.\eqref{eq:HAR}: with $\bm{X}$ being a vector of regressors, a general regression problem takes the form $Y = f\br{\bm{X},\bm{\beta}}$, with $f\br{\bm{X},\bm{\beta}} = \E\left[ Y \mid \bm{X} \right]$. Whether in the HAR model $f$ is a linear function of the parameters, in the CV-HAR setting, $f$ already embraces a conditional expectation and does not imply any strict structure (e.g. linear relationship between the regressors): $f\br{\bm{X},\bm{\beta}} = \E \left[\RKt | \RKCondThree \right]$.
In this light, the HAR model can be seen as a specific parametrization of a more general model which directly exploits conditional expectation $\E\left[\RKt | \RKCondThree \right]$, as our specification aims to.
Modelling the volatility at $t+1d$ as the expectation of a conditional distribution 
$F_{RK_{t+1d}^{\br{d}}|RK_t^{\br{d}}, RK_t^{\br{w}}, RK_t^{\br{m}}}$, by construction constraints the forecasts to the positive domain of $RK_{t+1d}^{\br{d}}$. Logarithmic specifications of the HAR model are no longer attractive, since this alternative approach circumvents positivity and normality issues, being naively suitable to cope with non-transformed realized measures. Furthermore, by modelling the joint distribution with copulas, transformations of the variables are not affecting the underlying copula \citep[][Theorem 2.4.3]{nelsen2007introduction}, so that also for modelling purposes transformations are irrelevant.
Also, the regression in eq.\eqref{eq:HAR} as such, produces symmetric confidence intervals based on t-distribution quantiles for the forecast values, while skewness and heavy-taildness are often observed in volatility.
On the contrary, the above approach leads to confidence intervals that are immediately identified by the actual quantiles of the same conditional distribution, potentially non-symmetric and showing kurtosis.
Since the joint distribution modeling based on C-Vine copulas presented in the following Section, we call such model CV-HAR\footnote{Where \enquote{CV} stands for \enquote{C-Vine}  and \enquote{HAR} is {\it nothing more than a reminder} of the motivation related to the HAR model. It does \textit{not} stand for \enquote{heterogeneous auto-regressive}.}.

\section{Modelling joint distributions with Vine copulas}\label{sec:Vines}
Although the wide range of flexible bivariate parametric copulas, the number of copula families available for multivariate (three or more variables) modeling is rather limited in contrast to the bivariate case. In the last two decades, a number of methods have been developed to construct high dimensional copulas with desirable proprieties \citep[see e.g.][]{joe2014dependence}.
Among them we mention: hierarchical (or nested) Archimedean copulas \citep[e.g][]{mai2012h}, mixtures of max-min infinitely divisible distributions \citep[e.g.][]{joe1996multivariate}, Factor copulas \citep[e.g.][]{hull2004valuation, mcneil2005quantitative,oh2017modeling} and Vine copulas \citep[e.g.][among the earliest works]{joe1994multivariate, joe1996families, cooke1997markov}. 
The pair construction method we adopt in this paper leads to the so-called Vine copulas (or Vines), see e.g. \citet{czado2010pair,joe2011dependence} for an introduction to Vine copulas. 

\subsection{Vine copulas}\label{sec:VineIntrod}
This Section aims at providing a short introduction to Vine copulas and in particular to the recursive pair-copulas construction method. Further details can be found e.g. in the monograph of \citet{joe2011dependence}. 
The starting point for constructing multivariate distributions is the well known recursive
decomposition of a multivariate density into products of conditional densities. Let $(X_1, ...,X_d)$ be a set of random variables with joint distribution $F$ and density $f$, let $F(\cdot|\cdot)$ and $f(\cdot|\cdot)$ denote conditional CDFs and densities respectively, then:
\begin{align}\label{eq:1Czado}
f\left( x_1,...,x_d \right)&=f\left( x_d|x_1,...,x_{d-1}\right)f\left(x_1,...,x_{d-1}\right)\\ \nonumber
&=f_1\left( x_1 \right) \prod_{i=2}^d f\left( x_i|x_1,...,x_{i-1}\right) \nonumber
\end{align}
As second ingredient we need Sklar's theorem (in its density form) to conveniently factorize a bivariate density $f (x_1,x_2)$ into a product of (unconditional) marginals $f_1$, $f_2$ and a bivariate copula density $c_{12}(\cdot,\cdot)$:
\begin{equation}\label{eq:2Czado}
f (x_1,x_2) = c_{12}(F_1(x_1),F_2(x_2))f_1(x_1)f_2(x_2)
\end{equation}
Using \eqref{eq:2Czado} we can express the
conditional density of $X_1$ given $X_2$ as
\begin{equation}\label{eq:3Czado}
f(x_1|x_2) = c_{12}(F_1(x_1),F_2(x_2))f_1(x_1)
\end{equation}
For distinct indices $i,j$, $i_1,...,i_k$ with $i<j$ and $i_1< ...<i_k$ we use the abbreviation:
$$
c_{ij|i_1,...,i_k} = c_{ij|i_1,...,i_k} \left( F(x_i|x_{i_1},...,x_{i_k} ),F(x_j |x_{i_1} ,...,x_{i_k} )\right)
$$
From equation \eqref{eq:3Czado} we can express $f\left( x_i|x_1,...,x_{i-1}\right)$ recursively, this yields to the expression:
\begin{equation}\label{eq:7Czado}
f\left( x_i|x_1,...,x_{i-1}\right)=c_{(i-1)i|1,...,i-2} \times f\left( x_i|x_i,...,x_{i-2}\right)
\end{equation}\label{eq:JoinFactorization}
Taking equation \eqref{eq:7Czado} in \eqref{eq:1Czado}, it follows that:
\begin{align}\label{eq:8Czado}
f\br{x_1,...,x_d} &= \prod_{j=1}^{d-1} \prod_{i=1}^{d-j} c_{j\br{j+i}|1,...,j-1} \times \prod_{k=1}^{d} f_k\br{x_k}\\ \nonumber
&=\prod_{j=1}^{d-1} \prod_{i=1}^{d-j} c_{j\br{j+i}|1,...,j-1} \br{ F\br{x_j|x_1,...,x_{j-1} },F\br{x_{j+i} |x_1 ,...,x_{j-1} } } \times \prod_{k=1}^{d}f_k\br{x_k}
\end{align}
This is called C-(canonical) Vine distribution.

Note that C-Vine decomposition of the joint density in eq.\eqref{eq:8Czado} consists of pair-copula densities $c_{ij|i_1,...,i_k}$ specified  for the variables indices $i,j$, conditioned to variables $i_1,...,i_k$, evaluated at the conditional CDFs  $F\left( x_i|x_{i_1},...,x_{i_k}\right)$, $F\left( x_j|x_{i_1},...,x_{i_k}\right)$ and marginal densities. This is why such a decomposition  is called pair-copula decomposition and the above construction leading to eq.\eqref{eq:8Czado} is called pair-copula construction. Importantly note that the decomposition is not unique, there are indeed $\frac{d\left(d-1\right)}{d}$ different sets of copulas to chose from and thus structures that build up to the joint distribution of variables $1,..,d$. Therefore, based on the specific problem under consideration, a specific tree needs to be identified, see Subsection \ref{subsec:tree_structure}.

\subsubsection{Estimation}\label{subsec:vineestim}
The standard framework for Vine estimation is likelihood maximization. From eq.\eqref{eq:8Czado} the log-likelihood $l$ is immediately recovered for a C-vine copula with parameter $\bm{\theta}_{CV}$, for a sample $\bm{u} = \br{u_{k,j}}$, with $k=1,...,N$ and $ j=1,...,d$:
\begin{equation}
l \br{\bm{\theta}_{CV}|\bm{u}} = \sum_{k=1}^{N} \sum_{i=1}^{d-1} \sum_{j=1}^{d-i} \log \left[ c_{i,i+j|1:\br{i-1}} \br{F_{i|1:\br{i-1}},F_{i+j|1:\br{i-1}}|\bm{\theta}_{i,i+j|1:\br{i-1}}} \right]
\end{equation}
where $F_{j|i_1:i_m} = F\br{u_{k,j}|u_{k,i_1},...,u_{k,i_m}}$ and the marginal distributions are uniform, i.e., $f_k\br{u_k} = \bm{1}_{\left[ 0,1 \right]}\br{u_k}$. $\bm{\theta}_{i,i+j|1:\br{i-1}}$ is the parameter set corresponding to the copula $c_{i,i+j|1:\br{i-1}}$. Note that according to eq.\eqref{eq:ConditionalCDFAll} $F_{j|i_1:i_m}$ depends on the parameters of pair-copula terms in tree 1 up to tree $i_m$ \citep{brechmann2013cdvine}. For extensions to general R-Vine structures see e.g. \citet{joe2011dependence}.

The likelihood maximization is not limited to the best parameter selection but defines the copula parametric specifications for the copulas in all the trees. While the unconditional copulas on tree 1 one can be easily specified by the researcher simply by using the input data, the copulas on the conditional CDF in higher trees are not directly available, since the conditional sample is not observed. 
Therefore the most common way for the parametric copula specification at trees $m>1$ is by recursively trying different copulas families for each conditional copula and retain the one leading to the highest likelihood in its fitted parameters. I.e. a given parametric copula $C_{i,i+j|1:\br{i-1}}$ is selected by recursively evaluating the likelihood of the data with different copulas specifications (Archimedeans, Gaussian and t- copulas). The Vine copula specification leading to the highest overall likelihood is then retained (actually the criterion we use is based on This copula selection proceeds tree by tree, since the conditional pairs in trees $2,...,m-1$ depend on the specification of the previous trees. 
Fig.\ref{fig:CVineStructure} (lower panel) provides an illustration of an estimated Vine.

It is a good practice to inspect the unconditional copulas specification at the first tree with alternative methods, since misspecifications would propagate through the whole tree and affect all the other parameters' estimates. These include graphical methods such as quantile dependence plots \cite{oh2017modeling},  contour plots for the fitted copula, $\chi$ and $k$ plots and likelihood ratio tests such as Voung and Clarke \citep{vuong1989likelihood,clarke2007simple}. See \citep{brechmann2013cdvine} and references therein for further details.

Pair copula construction therefore allows to sequentially combine specific pair copulas to build up a multivariate copula (and thus distribution) by identifying suitable bivariate pair-copulas:
\begin{itemize}
\item[i.] Use the sample to model the marginal distributions $F_i$, $i=1,...,d$ involved in the first tree. Use the margins to reduce the sample to the $\left[ 0,1 \right]$ interval.
\item [ii.] Specify the tree structure, i.e. type of vine and variable order, Subsection \ref{subsec:tree_structure})
\item[iii.] Determine conditional and unconditional copulas $c_{ij|i_1,...,i_k} $ families and parameters on the basis of the above discussion.
\item[iv.] Apply eq.\eqref{eq:8Czado} to build the multivariate distribution based on vine copula construction.
\end{itemize}  
Importantly, the above procedures can be exploited to test for independence by using the independent copula $C\br{u,v} = uv$. along with test statistics for the Spearman's rho  and Kendall's tau dependence measures \citep[see e.g.][for their asymptotics]{genest2007everything}.

\subsubsection{Tree-structure representation}\label{subsec:tree_structure}
As earlier mentioned, a density $f\left(x_1,...,x_d\right)$ can be represented by a product of pair-copula densities and marginal densities. The decomposition is however not unique. For instance, with $d = 3$, a possible decomposition for $f\left(x_1,x_2,x_3\right)$ is:
\begin{align*} \nonumber
    &f\left(x_1,x_2,x_3\right) =
     f\left(x_3|x_1,x_2\right)f\left(x_2|x_1\right)f_1\left(x_1\right) =\\ \nonumber
    & = c_{13|2}\left( F_{1|2}\left( x_1|x_2\right), F_{3|2}\left( x_3|x_2\right)\right) f_{3|2}\br{x_3|x_2} \times c_{12}\left( F_{1}\br{x_1}, F_{2}\br{x_2}\right) f_{2}\br{x_2} \times f_{1}\br{x_1} \\ \nonumber
    & = c_{13|2}\left( F_{1|2}\br{x_1|x_2}, F_{3|2}\br{x_3|x_2}\right) c_{23}\left( F_{2}\br{x_2}, F_{3}\br{x_3}\right) f_{3}\br{x_3} \times c_{12}\left( F_{1}\br{x_1}, F_{2}\br{x_2}\right) f_{2}\br{x_2} \times f_{1}\br{x_1}\\\nonumber
    & =f_{1}\br{x_1} f_{2}\br{x_2} f_{3}\br{x_3} \times c_{12}\left( F_{1}\br{x_1}, F_{2}\br{x_2}\right) \times c_{23}\left( F_{2}\br{x_2}, F_{3}\br{x_3}\right) c_{13|2}\left( F_{1|2}\br{x_1|x_2}, F_{3|2}\br{x_3|x_2}\right) \\ 
    & = f_1 f_2 f_3 \times c_{12} \times c_{23} \times c_{13|2}
\end{align*}
However by applying a different conditioning,
\begin{align*}
    f\left(x_1,x_2,x_3\right) &= f\left(x_1|x_2,x_3\right)f\left(x_2|x_3\right)f\left(x_3\right) 
    = ... = f_1 f_2 f_3 \times c_{13} \times c_{23} \times c_{12|3}
\end{align*}
which is clearly a different decomposition of $f\left(x_1,x_2,x_3\right)$.

\begin{figure}[ht!]
    \centering
    \includegraphics[scale = 0.50, angle = 0]{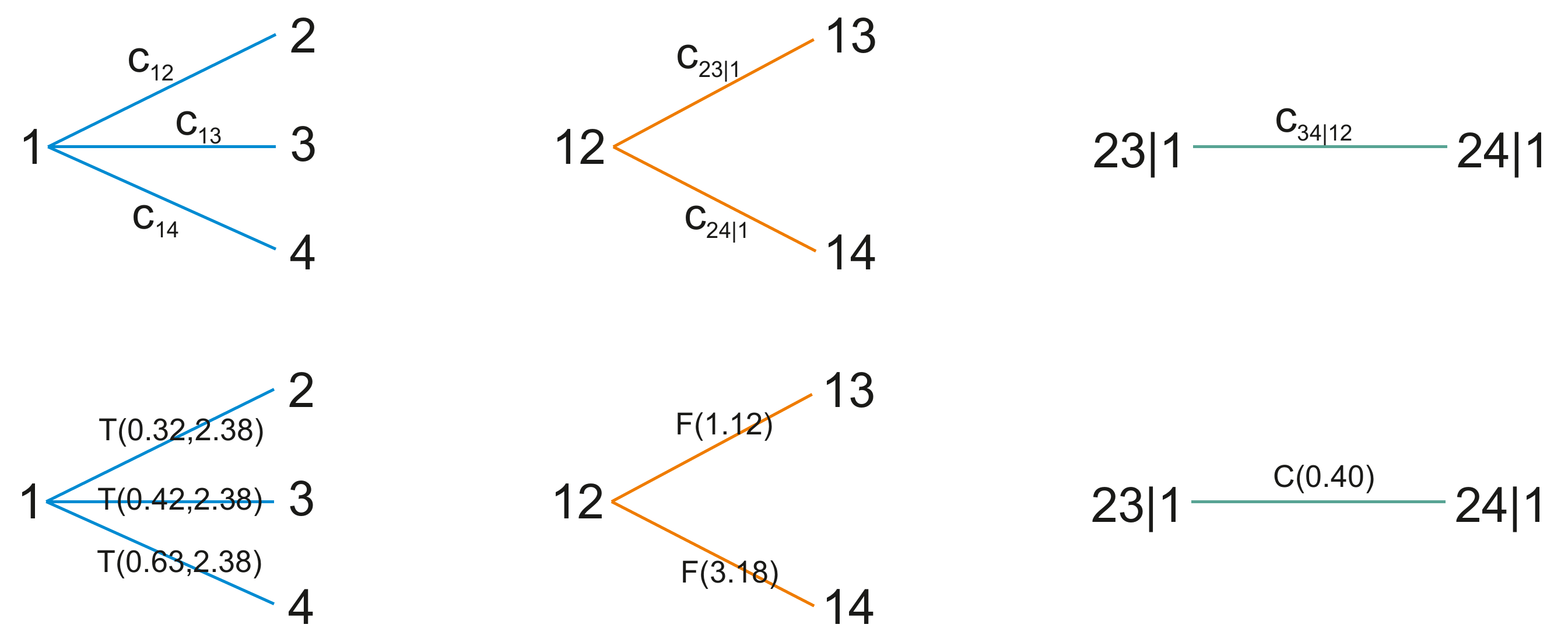}
    \caption{\textit{Upper panel}. C-Vine representation. This corresponds to the actual structure implemented in the application. Label \enquote{1} corresponds to monthly volatility, \enquote{3} to weekly volatility, \enquote{2} and \enquote{1} respectively to yesterday's and today's volatility terms. The tree structure corresponds to the following decomposition of the joint density: $f_{1234} = f_1\cdot f_2\cdot f_3\cdot f_4 \cdot{\color{NavyBlue} c_{12}} \cdot  {\color{NavyBlue} c_{13}}  \cdot {\color{NavyBlue}c_{14}} \cdot {\color{Orange} c_{23|1}} \cdot {\color{Orange} c_{24|1}} \cdot {\color{ForestGreen} c_{34|12}}$. \textit{Bottom panel}. Vine copula estimation example. Stock AXP, day 250, ECDF margins, estimation window: day 1 to 250. \enquote{T} stands for t-copula, \enquote{F} for Frank copula, \enquote{C} for Clayton copula. Numbers in the brackets are parameters' estimates.}
    \label{fig:CVineStructure}
\end{figure}

In general $d$-dimensional joint distributions allows for $d\left(d-1\right)/2$ different pair-copulas. \citet{bedford2001probability} introduced a tool called regular vine (R-Vine) structure to help to organized them. The formal definition of a regular vine they provide allows for a convenient graphical representation of the structure in terms of trees and nodes. Importantly, among the regular vines, an important sub-class is that of the so-called canonical vines (C-Vine). Fig.\ref{fig:CVineStructure} provides a graphical representation of a C-Vine. Canonical vines have a typical  \enquote{star} structure, where each tree has a unique node connected to all the other nodes. By defining the \textit{degree} of a node as the number of nodes attaching to it, for a $d$-dimensional problem, a C-Vine is a structure such that each node in tree $T_j$, $j=1,\dots d-1$ is of maximal degree, i.e. each tree $T_j$ has a unique node of degree $j-1$. 

Fig.\ref{fig:CVineStructure} clarifies the above statement. The C-vine density therein depicted, corresponds to the factorization;
$$
 f_{1234} = f_1 \cdot f_2 \cdot f_3 \cdot f_4 \cdot c_{12} \cdot  c_{13} \cdot c_{14} \cdot c_{23|1} \cdot c_{24|1} \cdot c_{34|12}
$$
where $X_1$ is set as a node in tree $T_1$, and the dependence with any other variable is considered with respect to it. I.e. the involved pair of variables are 12, 13, 14 with their respective pair-copulas $c_{12}$, $c_{13}$, $c_{14}$. In three $T_2$, the connection involves $F_{12}$, $F_{13}$, $F_{14}$, with $F_{12}$ as node. Conditional on the common variable (in the Fig.\ref{fig:CVineStructure} always 1, corresponding to $X_1$) the dependence between $F_{2|1}$ and $F_{3|1}$ is captured by the copula $c_{23|1}$ (which would naturally arise given a proper factorization of $f\br{x_1,.x_1,x_2,x_4}$ by applying the law of total probability, similarly as eq.\eqref{eq:JoinFactorization}).
Analogously $c_{24|1}$ connects $F_{2|1}$ and $F_{4|1}$. In the last tree $T_3$ a path connecting 23\textbar1 and 24\textbar1 ($F\br{X_2|X_1,X_3|X_1}$ and $F\br{X_2|X_1,X_4|X_1}$) is given by the copula $c_{23|12}$. Coherently with the above definition, C-Vines always show a path connection (rather than a star) in the last tree $T_{d-1}$.

C-Vines log-likelihood has the convenient form of eq.\eqref{eq:8Czado}, that is of immediate evaluation given the result discussed in Section \ref{subsec:Condexpectations} about the computation of conditional CDFs. 
About the structure selection, some guidelines are provided e.g. in \cite{dissmann2013selecting,czado2013selection}. The intuition of selecting the structure leading to maximum likelihood is unfeasible for-large dimensional problems. Some hypothesis on the structure, i.e. on the relationship between variables must be accounted for in order to simplify the selection problem. 
Thus the nature of the problem and its interpretability are important drivers in structure selection. For the volatility modeling problem here analyze we chose a C-Vine structure, although other structures might be applicable and of feasible interpretation too. See section \ref{subsec:VineCopConstr} for further details in this regard.

\subsection{Conditional distributions and expectations from Vine copulas}\label{subsec:CondDistrCopulas}
In our CV-HAR specification the quantity of interest is an expectation having form $\E \left[ X_1|X_2,X_3,X_4 \right ]$. Here we discuss how such an expectation is obtained from the conditional distribution extracted from the overall Vine joint.
\subsubsection{Conditional distribution}
By now, consider the simplest case of $X_1$ and $X_2$ being uniformly distributed, $C$ is their copula and $Y$ their joint CDF. Be $0 \leq \epsilon \leq 1- x_2$, with $x_1, x_2 \in \mathbb{R}$.
\begin{align*}
    \Prob\br{X_1 \leq x_1, X_2 \in \br{x_2, x_2 + \epsilon}} &= \Prob\br{X_1 \leq x_1,X_2 \leq x_2+\epsilon}-\Prob\br{X_1 \leq x_1,X_2 \leq x_2}\\
    &= Y\br{x_1,x_2 +\epsilon} - Y\br{x_1,x_2}\\
    &= C\br{x_1,x_2 + \epsilon} - C\br{x_1,x_2}
\end{align*}
Since $X_2$ is uniform, $\Prob\br{X_2 \in \br{x_2, x_2 + \epsilon}} = \epsilon$, by Bayes the theorem we compute the conditional probability:
\begin{equation*}
 \Prob\br{X_1 \leq x_1|X_2 \in \br{x_2, x_2 + \epsilon}} = \frac{C\br{x_1,x_2 + \epsilon} - C\br{x_1,x_2}}{\epsilon}  
\end{equation*}
By letting $\epsilon \rightarrow 0$ (provided that the limit exists) the partial derivative arises and the conditional CDF $X_1|X_2 = x_2$ solves to \citep{nelsen2007introduction}:
\begin{equation*}
\Prob\br{X_1 \leq x_1 | X_2 = x_2} = \frac{\partial C\br{x_1,x_2}}{ \partial x_2}
\end{equation*}
The conditional CDF $F\br{x_1|x_2}$ turns to have an immediate expression in terms of the copula $C$ between $X_1$ and $X_2$: just take its partial derivative with respect to the conditioning variable and evaluate it in $\br{x_1,x_2}$.

The case with $X_1$ and $X_2$ distributed according to $F_1$ and $F_2$ (in general different and not uniform), by Sklar's theorem, is immediately solved by updating the arguments in which the copula is evaluated:
\begin{equation}\label{eq:SimpleBivDecompos}
F\br{x_1|x_2} = \Prob\br{X_1 \leq x_1 | X_2 = x_2} = \frac{\partial C\br{F_1\br{x_1},F_2\br{x_2}}}{ \partial F_2\br{x_2}}
\end{equation}
Similar results can be proved for the conditional PDF, instead of CDFs.
By the same reasoning, expanding for $d = 3$, e.g. $F_{3|12}$ can be evaluated as follows: 
\begin{align*}
F_{3|12} &= \frac{\partial C_{32|1}\br{F_{3|1},F_{2|1}}}{\partial F_{2|1}}
= \frac{\partial C_{32|1} \br{\frac{\partial C_{13}\br{F_3\br{x_3},F_1\br{x_1}}}{ \partial F_1\br{x_1}},\frac{\partial C_{12}\br{F_2\br{x_2},F_1\br{x_1}}}{ \partial F_1\br{x_1}}}}{{\partial F_{2|1}} }
\end{align*}
where $C_{12}$ is the copula associated with the pair $\br{X_1,X_2}$, $C_{13}$ the copula for $\br{X_1,X_3}$ and $C_{32|1}$ the copula for $\br{X_3|X_1,X_2|X_1}$.
It clearly emerges that the earlier pair copulas $C_{12}$ and $C_{13}$ are sequentially used to estimate the conditional CDFs. And that these constitute the arguments of the higher-order conditional copula $C_{32|1}$, from which the higher-order conditional CDF $F_{3|12}$ is constructed.

In a general setting \citet{joe1996families} obtains the following recursive relationship, for $F\br{x|\bm{v}}$:
\begin{equation} \label{eq:ConditionalCDFAll}
    F\br{x|\bm{v}} = \frac{\partial C_{xv_j|\bm{v}_{-j}}\br{F\br{x| \bm{v}_{-j}},F\br{v_j| \bm{v}_{-j}}}}{\partial F\br{v_j| \bm{v}_{-j}}}
\end{equation}
where $\bm{v}$ is a $m$-dimensional vector, $v_j$ any arbitrary component of $\bm{v}$ and $\bm{v}_{-j}$ the $(m-1)$-dimensional vector obtained by excluding $v_j$ from $\bm{v}$. Importantly, note that $C_{xv_j|\bm{v}_{-j}}$ is always a bivariate copula function. \\
This is a crucial result for the conditional CDF construction, since it shows that $F\br{x|\bm{v}}$ can be obtained by {\it sequentially mixing} conditional CDFs with copulas, where the conditional copula $C_{xv_j|\bm{v}_{-j}}$ depends on the copulas $C_{xv_i|\bm{v}_{-ij}}$ and $C_{v_j v_i|\bm{v}_{-ij}}$, conditional on the smaller set $\bm{v}_{-ij}$, and so backwards up to the unconditional copulas on the first tree.

It is clear how Vine copulas are particularly attractive in terms of the recursive relation in eq.\eqref{eq:ConditionalCDFAll}. Once the structure is specified, all the conditional copulas are settled in terms of their parametric characterization. All the conditional distributions $F\br{x|\bm{v}}$, recursively determined by the partial derivatives of the copulas identified in the previous tree, are determined as well. Once the vine structure is estimated, all the pair-copula parameters are known and the Vine is completely determined. By simple substitution of the conditioning values the variables' CDFs, $F\br{x|\bm{v}}$ is readily computed.

\subsubsection{Conditional expectation}\label{subsec:Condexpectations}
The conditional CDF can be recursively evaluated given the convenient Vine decomposition.
This stands as a starting point to evaluate conditional expectations of the type $\E \left[ X|\bm{v} \right ]$. Solving $\E \left[ x|\bm{v} \right ]$ is in general an integration problem, whose complexity depends on the parametric copulas involved in eq.\eqref{eq:ConditionalCDFAll}.
In this research, we follow a non simulation-based approach for computing the conditional expectation \citep[improving the procedure of][where however a much simpler bivariate problem is considered]{sokolinskiy2011forecasting}.
Although the most used definition for computing the expectation is in terms of integral with respect to the conditional PDF ($f\br{x|\bm{v}}$), eq.\eqref{eq:ConditionalCDFAll} involves CDFs.
Not to differentiate twice the CDF to extract the corresponding PDF, but to use eq.\eqref{eq:ConditionalCDFAll} directly we adopt the following alternative in terms of CDF:
\begin{equation} \label{eq:CDFIntegration}
    E\left[ X| \bm{v} \right] = -\int_{-\infty}^{0} F\br{u|\bm{v}} du + \int_{0}^{\infty } 1-F\br{u|\bm{v}} du
\end{equation}
The joint implementation of the nested structure and integration  of eq.\eqref{eq:ConditionalCDFAll} and eq.\eqref{eq:CDFIntegration} is complex. We verify the above implementation by comparing conditional expectation computed via simulation for a number of different Vines.

\section{Empirical application}\label{sec:EmpApp}
\subsection{Data}
This research uses trade data for 10 of the 30 stocks constituting the Dow Jones industrial average index. The stocks under consideration in this analysis are, AAPL, AXP, BA, CAT, CSCO, CVX, DIS, GS, HD, IBM. In order to have a data-sample large enough for in-sample and out-of-sample analyses, the data covers a long span of 1634 days\footnote{The length of the data used in the application reduces to 1626 days, by excluding 22 days for constructing the first $\RKw$.}, from January 1\textsuperscript{st} 2012 to June  30\textsuperscript{th} 2018. The data is extracted from the TAQ database, consisting of raw trade prices, their respective timestamps, quantities and other fields identifying e.g. the exchange. For each stock raw prices have been preprocessed and cleaned according to the guidelines presented in \citet[][Section 3.1]{barndorff2009realized}. To avoid biases induced by non-regular trading hours entries outside 9:30am-4pm have been removed, as well as entries with transaction price equal to zero. For the 10 stocks the exchanges on which the trading activity took place have been recorded, and their absolute frequencies analyzed. We retain the data from the exchange \enquote{D} - Financial Industry Regulatory Authority, Inc. (FINRA ADF)\footnote{TAQ manual: Daily TAQ client specification. Version 2.2a.}- it collects 25.13\% of the total number of transactions that occurred in the period analyzed (and traded all the above 10 stocks, for every day).
Entries with abnormal sale condition (not corrected or canceled by the participant), were also removed. Multiple transactions sharing the same timestamp have been replaced by their median price. The timestamp resolution varied in the period under investigation, however, the accuracy is up a millisecond. Entries for which the transaction price mean absolute deviation deviated by more than 10 times the average MAD computed over a centered window of 2 minutes length., have been removed as well\footnote{This partially resembles the rule \enquote{Q4} of \citet{barndorff2009realized}: besides applying the previous cleaning steps, occasionally there are extreme outliers, visually not coherent with the average daily behavior of the price series.}. By averages over the last 5 and 22 RK measures, we compute $\RKw$ and $\RKm$.

\subsubsection{Margins}
Modelling the CDFs corresponding to the four\footnote{Note that unconditionally $RK_{t+1d}^{\br{d}}$ and $RK_{t}^{\br{d}}$ share the same distribution. In practice, we deal with tree marginal distributions.} volatility terms is a sensible step. On first instance this leads to the transformed sample in $\left[0,1 \right]$ interval upon which the copula model is built, secondly it drives the implementation of eq.\eqref{eq:ConditionalCDFAll} and eq.\eqref{eq:CDFIntegration}. Not to obtain results that are specific and valid under a given procedure for computing the CDFs, for each of the four volatility components $\RKCondFour$ we perform the analysis by the use of (i) parametric CDFs, (ii) Kernel-based CDFs (estimated over positive domains) and (iii) ECDFs. In the parametric case we fit an Inverse-Gaussian (IG) distribution, motivated by the arguments of \citep[e.g.][]{barndorff2002econometric,forsberg2002bridging}.

\subsubsection{Vine-Copula construction}\label{subsec:VineCopConstr}
The C-Vine copula construction follows the methodology described in Section \ref{sec:Vines}. The copula models considered in the analyses are the Archimedean copulas (Gumbel, Frank, Joe, Clayton), the Gaussian copula, and the t-copula \citep{joe2014dependence}. The implementation of eq.\eqref{eq:ConditionalCDFAll} does not pose problems for the Archimedean copulas (all the copulas are smooth functions), while the differentiation of the Gaussian and particularly the t-copula is achieved numerically with the approximation $\partial C\br{u,v}/\partial v = \br{C\br{u,v+h}-C\br{u,v}}/h$ with $h = 0.001$. The use of Gaussian and t-copula constitutes a computational complication, which is however not granted to lead to a considerable gain in fitting performance and, in the latter, forecast improvements with respect to the Archimedean set. Therefore the analyses are conducted separately, for Vine constructions estimated using Archimedeans copulas only (\enquote{A} set) and for Vines allowing for of all the six alternatives  (\enquote{AGT} set).

The copula selection in the different trees is automated based on the AIC criterion. However, misspecifications in the first tree would propagate through the whole Vine structure. Therefore, for a selected number of stocks and a range of days, we double-check by use of the alternative selection methods described Section \ref{subsec:vineestim}.

As mentioned in Section \ref{sec:Vines}, the choice of the structure is not unique. We use a C-Vine copula representation.
The C-Vine structure is built in such a way that $\RKm$ constitutes a node on the first tree. This is a feasible choice, assuming a cascade effect of past volatilities to the current one, i.e. that the past volatility and its history drive today's - modeling in the first tree last moth's volatility given today's is not logical as modeling today's given last month's. Note that the intuition of using $\RKt$ as a node on the first tree is not feasible since our target is the estimation of $\E\left[\RKt | \RKCondThree \right]$ by eq.\eqref{eq:ConditionalCDFAll}: within this framework $RK_t$ can only by conditioned and not conditioning. In this way,  the first tree considers all the dependencies between the volatility variables and $\RKm$ . On the last tree, in the conditional copula $c_{\RKt \RKy | \RKw \RKm}$ today's and yesterday's volatility are conditioned to last week's and month's, which is coherent and intuitive from a logical point of view (see Fig.\ref{fig:CVineStructure}, upper panel). The estimation follows Section \ref{subsec:vineestim}. The conditional CDF $F_{\RKt | \RKCondThree}$ is retrieved by applying eq.\eqref{eq:ConditionalCDFAll}, whereas the conditional expectation $\E \left[ \RKt \vert  \RKy = x_t^{\br{d}}, \, \RKw = x_t^{\br{w}}, \, \RKm = x_t^{\br{m}} \right]$ is evaluated with eq.\eqref{eq:CDFIntegration}. 

\subsection{A simple non-linear benchmark model}
As a benchmark for comparing the CV-HAR model, we implement a simple neural network (NN) model. 
We follow the methodological approach of \cite{arneric2018neural}, by implementing a feed-forward neural network with a single hidden layer embracing two neurons. \cite{arneric2018neural} argues that such a network design is optimal in terms of in-sample MSE, prevents over-fitting and is parsimonious. A logistic activation function is applied between input and hidden layers, and a linear function between the hidden and output layers \citep[e.g.][]{medeiros2006building}. The very same regressors as for the HAR and CV-HAR models are used for the NN estimation, namely $\RKy$, $\RKw$, $\RKm$. Following \citep{medeiros2006building,hillebrand2010benefits}, the network is estimated via Bayesian regularization \citep{mackay1992practical}, along with the Levenberg-Marquardt optimization algorithm \citep{hagan1994training}. 70\% of the data is used as a training sample, the remaining 30\% for validation. Splits are randomly initialized, as well as the initial parameters. Hence, each of the NN forecasts is computed by averaging over 500 bootstraps.

\section{Results}\label{sec:Results}

\subsection{Estimation and forecasting schemes}
\begin{figure}[htbp]
    \centering
    \includegraphics[scale = 0.65]{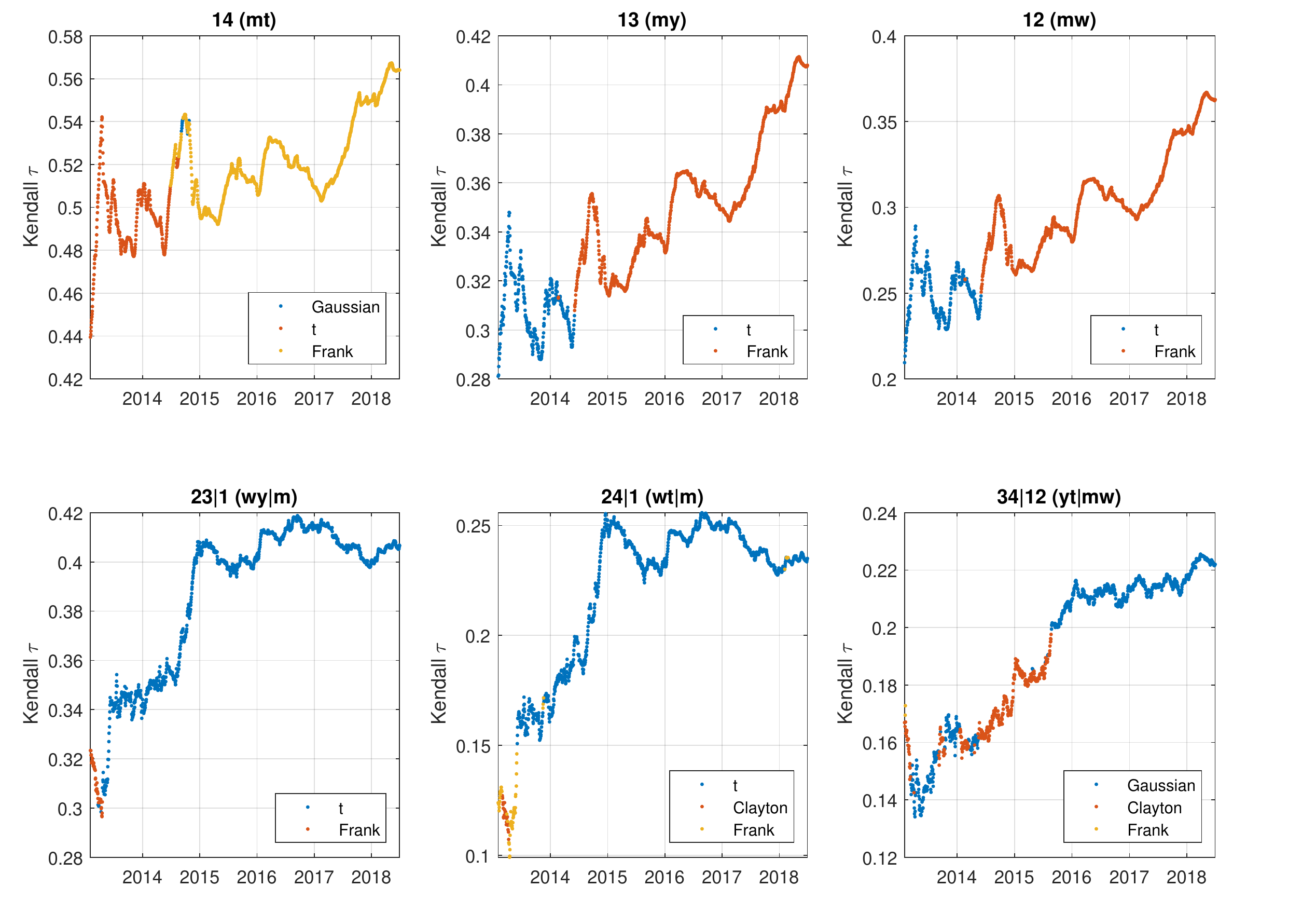}
    \caption{Selected Pair-copulas and implied Kendall's $\tau$ from the estimated parameters. The dependence modelled is identified in plots titles according to the notation in Section \ref{sec:Vines}. Variables 1, 2, 3, 4 respectively represent monthly volatility, weekly volatility, day-$t$ volatility, and day-$\br{t+1d}$ volatility components. AXP on day 250, ECDF margins and IW estimation.}
    \label{fig:RW500_kendall}
\end{figure}

The estimation of both the HAR and CV-HAR models, and thus the corresponding measures of forecast accuracy, are developed under three different schemes.
\begin{itemize}
    \item[i.] Fixed window (FW). Estimation of the models at days 250, 500, 750 and 1250. Thus, we estimate the models by using the first $W=\left\lbrace 250,500,750,1250\right\rbrace$ observations respectively. $W$ splits the sample in two. The part involving observations from day one to $W$, upon which the model is estimated, is used for in-sample analysis.
    The remaining part, from day $W+1$ to the last, is used for out-of-sample analysis. Coherently,  in-sample days from day one to $W$, constitute the training-validation set of the NN implementation (with a 70\%-30\% split). The remaining out-of-sample days, constitute the training set.
    \item[ii.] Increasing window (IW). We first estimate the HAR and CV-HAR models on days 1 to $W$, then sequentially for each day $d=W+i$, $ i=1 \dots \br{1626-1}-W$ we re-estimate the model by using the whole dataset up to day $d$. 
    With this procedure, for any day between $W+1$ and 1625, one-step-ahead forecasts are constructed, and the effect of sequentially increasing the sample size analyzed. Results are based on the following sizes of the first window: $W=\left\lbrace 250,500,750 \right\rbrace$.
    \item[iii.] Rolling window (RW). Similarly, as in (ii), we re-calibrate the model and construct one-step-ahead forecasts by using rolling windows of size $W=\left\lbrace 250, 500, 750\right\rbrace$ days. I.e. forecasts at day $d+1$ are based on models estimates with observations from $d-W+1$ to $d$, for $W+1\leq d \leq 1625$.
\end{itemize}
These three approaches provide different copula estimates. The IW and RW schemes allow for time-variation of the copula and therefore of the dependence between the pair-variables in the underlying copulas.
For the pair-copulas involved in the C-Vine tree, in Fig.\ref{fig:RW500_kendall} we show the estimated copulas with the increasing window approach. 
Since the parameters' space for different copulas is different, we plot the corresponding Kendall's-$\tau$ implied from the estimated copula.
Fig.\ref{fig:RW500_kendall} uncovers a time-varying nature of the dependence between the variables' pairs that the FW approach is unable to capture. Therefore, although IW and RW are much more demanding than FW from a computational point of view, including IW and RW provides a deeper insight into the complex dependence dynamics between the variables.

In the results' tables, we adopt the following measures to compare the performance of the CV-HAR against the HAR model. (i) Mean squared error (MSE), (ii) mean absolute error (MAE), (iii) median absolute deviation (MAD), (iv) mean absolute scaled error (MASE) \citep{hyndman2006another}, (v) mean absolute percentage error (MAPE), (vi) mean directional accuracy (MDA) and (vii) Qlik  \citep[e.g.][]{patton2009evaluating} which has been extensively used in similar applications \citep[see e.g.][]{patton2015good,bollerslev2016exploiting}\footnote{ With $\lbrace y_t \rbrace$ being sample values and $ \lbrace \hat{y}_t \rbrace $ their respective forecasts, with $i = 1,...,T$: $MASE = \frac{1}{T} \sum_{t=1}^T \frac{\mid \hat{y}_t - y_t \mid }{\frac{1}{T-1}\sum_{t=2}^T \mid y_t - y_{t-1} \mid}$, $MAPE = \frac{1}{T} \sum_{t=1}^T \mid \frac{y_t-\hat{y}_t}{y_t} \mid$, $ MDA = \frac{1}{T} \sum_{t=1}^T \bm{1}_{sign\br{y_t-y_{t-1}} == sign \br{\hat{y}_t - y_{t-1}}}$, $Qlik = \frac{y_t}{\hat{y}_t}-\log\br{\frac{y_t}{\hat{y}_t}}-1$.}. Results report the average measures over the 10 stocks. Whereas MSE, MAE, MAD, MAPE and Qlik are actual overall measures, MDA and MASE are averages of the 10 individual measures computed for each stock\footnote{MDA and MASE involve lagged values: it is not possible to compute them on an overall basis from an general time-series constructed by stacking the individual ones.}.

One-step-ahead forecasts of the two models have been tested to be statistically different with the Diebold-Mariano (DM) test \citep{diebold2002comparing} and the conditional predictive ability (CPA) test \citep{giacomini2006tests}. The well-known DM test applies to non-nested models only. For the IW scheme, the CPA test constitutes a suitable alternative. Under RW we apply both the DM and CPA tests. Tests are applied to squared, absolute and Qlik loss functions. This corresponds to test for differences in one-step-ahead MSE, MAE, and Qlik for the two models.
For the FW case, analyses are separate for the sample up to $W$, and from $W$ to 1625. These correspond to in-sample and out-of-sample analyses. For the IW and RW cases, measures refer to one-step-ahead forecasts. Accordingly, note that the measures are interpreted differently, e.g. MSE under FW is the in-sample MSE, while under RW is the one-step-ahead out-of-sample MSE.

Common NNs, and more generally machine-learning implementations are difficult to scale over RW and IW schemes. The implementation of the NN for the RW and IW schemes is computationally challenging and very demanding, if not unfeasible. E.g. just for the IW case with $W=250$ this would require $6.88\cdot 10^5$ estimations over an increasing data-sample (1363 forecasting days and 500 bootstraps) to average out the effects of the initial random sample split and weights. Such a complex and time-consuming estimation is out-of-scope wrt. the objectives and motivation of the present research: the NN is discussed under the FW scheme only. Broader NN applications and analyses are left for future research.

\subsection{In-sample analysis}
We present the results relative to the estimation of the CV-HAR model against the HAR model. Epochs at which the models are estimated under a fixed-window approach, corresponding to the width $W$ of the estimation period, are $W=\lbrace 250, 500,750,1250\rbrace$. 
This analysis focuses on the in-sample accuracy of the two models under investigation. Since this does not involve any (one-step-)ahead forecasting, no formal testing of forecasting accuracy is here developed.
Results in tab.\ref{tab:FW_ISOS} show that on a general level the window size has an important impact. Indeed, individual measures are generally improving as the size of the window widens. In this regard, under $W=250$ we do not observe a uniform improvement of the CV-HAR model over the HAR, which is remarkable for wider windows. However, at the shortest window $W=250$,  improvements over the HAR model are observed corresponding to copula constructions allowing for Gaussian and t-copulas besides Archimedeans. This indicates that in small samples, flexibility on pair-copulas leads to clear improvements over a constrained framework allowing for Archimedeans only. This pattern applies in general too, i.e. measures corresponding to constructions relying on Archimedeans only are outperformed by those extending the set to Gaussian copula and t-copula too. This is however not crucial under wider windows, where the model is always satisfactory, suggesting that the role of the extended set at short window lengths is that of correcting for small-size distortions in the joint distribution of the volatility terms captured by the C-Vine copula. On the other hand, models for margins seems not to have a role in this analysis, since all the constructions lead to similar ratios.
Furthermore, note that the mean directional accuracy (MDA) is generally close to unity. This indicates that the HAR and CV-HAR models are equally capable of forecasting the direction of tomorrows' volatility movement wrt. to today's (i.e. increase or decrease).

\begin{figure}[hb!]
    \centering
    \includegraphics[scale = 0.65, angle = 0]{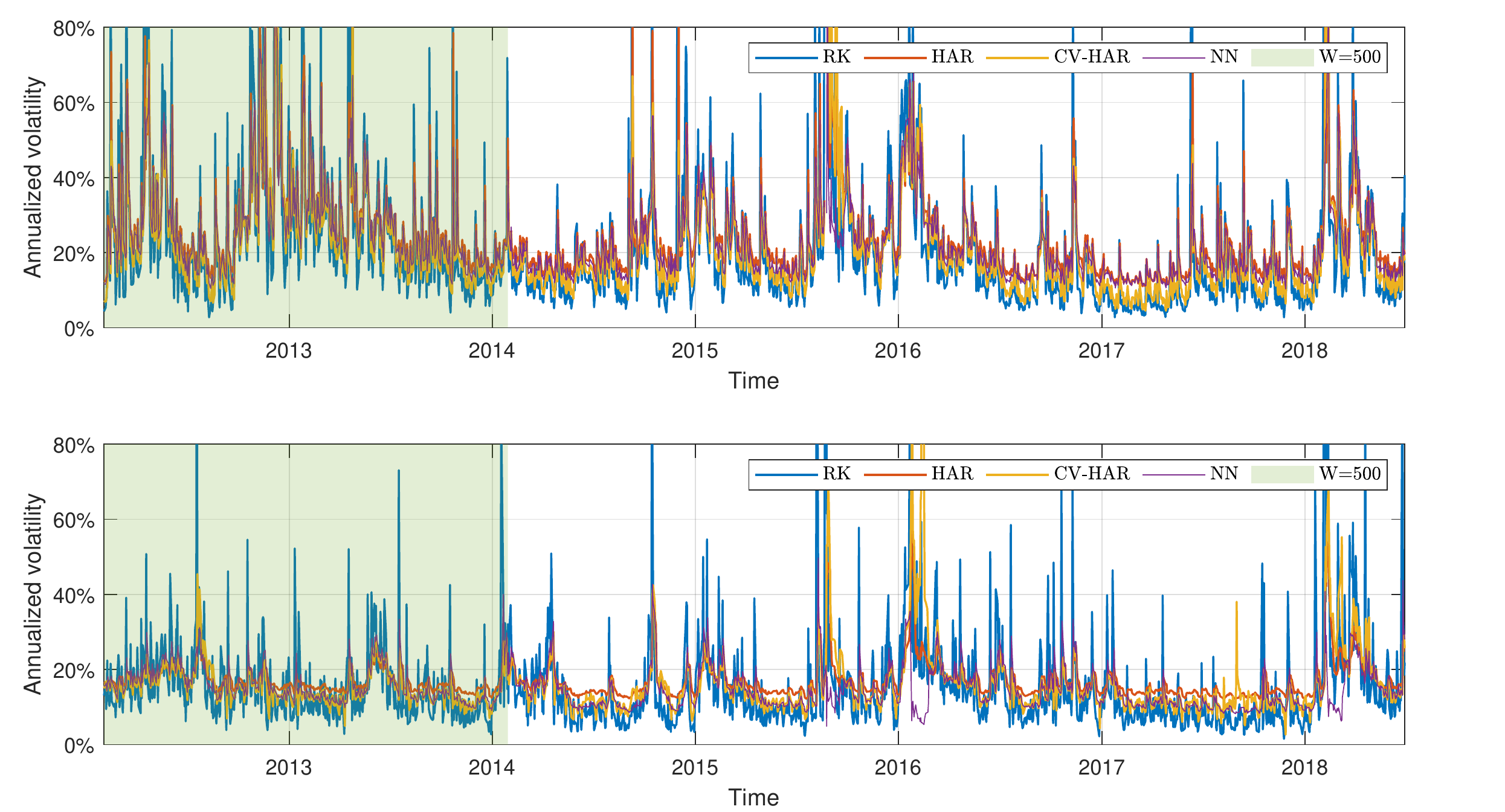}
    \caption{Sample series for AAPL and AXP. C-Vine construction over ECDF margins and Archimedeans, Gaussian, t- pair-copulas. Fixed window estimation with $W=500$. }
    \label{fig:2Series}
\end{figure}

Overall, on an in-sample basis, these results seem to favor the CV-HAR flexibility allowing for conditional means of generic non-linear nature. The linear combination between the volatility components of the HAR model, besides being outperformed for most of the measures, is not guaranteed to produce positive volatility forecasts. As it appears from Fig.\ref{fig:2Series}, HAR estimates are in general around an average volatility level, not accurately tracking neither days of low nor high volatility. In fact, a simple linear model in the three volatility terms, without any further dummies,  approximates today's volatility overall average behavior. On the other hand, the CV-HAR seems capable of reacting, although less promptly than the HAR model, to high volatility periods by generating appropriate estimates and non-linearly adapting to low the different regimes observed in the sample. 

That the CV-HAR successfully identifies a non-linear pattern is confirmed by the vicinity of the performance measures wrt. to those from the NN. On an in-sample basis, the NN seems to outperform the CV-HAR model wrt. MSE measure. This is not surprising considering that the NN estimation explicitly seeks for the set of parameters minimizing the MSE. On the contrary, the CV-HAR model is based on the apparently unrelated copula fitting via likelihood, from which forecasts are indirectly extracted.  Because of the completely different modeling and underlying assumptions, construction, and estimation approaches, and because of the very-different algorithm complexities, is remarkable, the CV-HAR setting leads to comparable results wrt. the NN implementation.
The out-of-sample panel of Tab.\ref{tab:OS}, on the contrary, is not favoring the NN. On the test-set over which the NN has not been optimized, its actual forecasting performance emerges. The non-linear regression functional that the CV-HAR model approximates seems to lead to better performance measures wrt. the NN, whose approximation appears to be constrained on the training set, not generalizing on new data. This is as stronger as the out-of-sample window size $W$ grows. The growing information conveyed in the joint distribution that the CV-HAR model exploits seem to be valuable in predicting future's RK dynamics.

\begin{table}[ht!]
  \centering
  \scalebox{0.75}{
    \begin{tabular}{lccccccc|ccccccc}
          & \multicolumn{7}{c}{In-sample}                         & \multicolumn{7}{c}{Out-of-sample} \\
          & MSE   & MAE   & MAD   & MASE  & MAPE  & QLIK  & MDA   & MSE   & MAE   & MAD   & MASE  & MAPE  & QLIK  & MDA \\
    \midrule
    $\mathbf{W=250}$ &       &       &       &       &       &       &       &       &       &       &       &       &       &  \\
    $HAR$ & \textit{0.672} & \textit{0.494} & \textit{0.334} & \textit{0.839} & \textit{0.437} & \textit{0.113} & \textit{0.668} & \textit{1.190} & \textit{0.534} & \textit{0.355} & \textit{0.992} & \textit{0.567} & \textit{0.113} & \textit{0.622} \B \\
    $NN$  & 0.988 & 0.994 & 1.004 & 0.994 & 0.995 & 0.985 & 0.995 & 1.112 & 1.037 & 1.016 & 1.051 & 1.007 & 0.985 & 1.007  \\ 
    $E_{A}$ & 1.035 & 1.042 & 1.068 & 1.071 & 1.086 & 1.047 & 1.040 & 1.065 & 0.996 & 0.951 & 1.010 & 0.939 & 1.047 & 1.046 \\
    $E_{AGT}$ & 1.019 & 0.981 & 0.934 & 0.991 & 0.951 & 1.024 & 1.001 & 1.093 & 0.976 & 0.899 & 0.991 & 0.936 & 1.024 & 1.012 \\
    $K_{A}$ & 1.030 & 1.031 & 1.060 & 1.059 & 1.071 & 1.044 & 1.037 & 1.062 & 0.985 & 0.942 & 1.000 & 0.924 & 1.044 & 1.038 \\
    $K_{AGT}$ & 1.022 & 0.977 & 0.935 & 0.986 & 0.945 & 1.015 & 1.004 & 1.002 & 0.954 & 0.899 & 0.965 & 0.903 & 1.015 & 1.015 \\
    $P_{A}$ & 1.027 & 1.018 & 1.027 & 1.030 & 1.046 & 1.024 & 1.032 & 1.035 & 0.993 & 0.959 & 1.003 & 0.954 & 1.024 & 1.028 \\
    $P_{AGT}$ & 1.016 & 0.988 & 0.979 & 1.000 & 0.990 & 1.012 & 1.011 & 0.993 & 0.972 & 0.934 & 0.981 & 0.940 & 1.012 & 1.018 \\
    \midrule
    $\mathbf{W=500}$ &       &       &       &       &       &       &       &       &       &       &       &       &       &  \\
    $HAR$ & \textit{0.690} & \textit{0.471} & \textit{0.326} & \textit{0.862} & \textit{0.465} & \textit{0.121} & \textit{0.660} & \textit{1.302} & \textit{0.554} & \textit{0.364} & \textit{1.017} & \textit{0.583} & \textit{0.121} & \textit{0.613} \B \\
    $NN$  & 0.957 & 0.957 & 0.933 & 0.961 & 0.935 & 0.961 & 0.992 & 1.148 & 1.009 & 0.915 & 1.027 & 0.916 & 0.961 & 1.001 \\
    $E_{A}$ & 0.997 & 1.016 & 1.044 & 1.032 & 1.040 & 1.007 & 1.042 & 1.055 & 0.964 & 0.898 & 0.976 & 0.893 & 1.007 & 1.015 \\
    $E_{AGT}$ & 0.981 & 0.956 & 0.917 & 0.970 & 0.919 & 0.987 & 0.996 & 1.049 & 0.966 & 0.872 & 0.983 & 0.917 & 0.987 & 1.006 \\
    $K_{A}$ & 0.995 & 1.007 & 1.028 & 1.024 & 1.022 & 1.002 & 1.040 & 1.029 & 0.957 & 0.908 & 0.974 & 0.894 & 1.002 & 1.015 \\
    $K_{AGT}$ & 0.978 & 0.952 & 0.912 & 0.962 & 0.912 & 0.979 & 0.997 & 1.001 & 0.948 & 0.865 & 0.962 & 0.885 & 0.979 & 1.003 \\
    $P_{A}$ & 0.999 & 1.037 & 1.104 & 1.062 & 1.096 & 1.026 & 1.053 & 1.089 & 0.994 & 0.957 & 1.010 & 0.940 & 1.026 & 1.025 \\
    $P_{AGT}$ & 0.982 & 0.979 & 0.984 & 0.992 & 0.980 & 0.987 & 1.013 & 1.013 & 0.974 & 0.918 & 0.986 & 0.929 & 0.987 & 1.017 \\
    \midrule
    $\mathbf{W=750}$ &       &       &       &       &       &       &       &       &       &       &       &       &       &  \\
    $HAR$ & \textit{0.747} & \textit{0.475} & \textit{0.332} & \textit{0.914} & \textit{0.508} & \textit{0.129} & \textit{0.642} & \textit{1.416} & \textit{0.573} & \textit{0.369} & \textit{1.013} & \textit{0.580} & \textit{0.129} & \textit{0.617} \B \\
    $NN$  & 0.869 & 0.937 & 0.875 & 0.942 & 0.891 & 0.926 & 0.990 & 1.976 & 1.139 & 0.848 & 1.143 & 0.921 & 0.926 & 0.994 \\
    $E_{A}$ & 0.977 & 0.971 & 0.930 & 0.981 & 0.920 & 0.943 & 1.029 & 1.027 & 0.930 & 0.824 & 0.944 & 0.822 & 0.943 & 1.016 \\
    $E_{AGT}$ & 0.965 & 0.949 & 0.872 & 0.963 & 0.886 & 0.929 & 1.007 & 0.958 & 0.904 & 0.784 & 0.919 & 0.789 & 0.929 & 1.004 \\
    $K_{A}$ & 0.981 & 0.969 & 0.939 & 0.982 & 0.921 & 0.949 & 1.030 & 0.988 & 0.925 & 0.845 & 0.936 & 0.826 & 0.949 & 1.021 \\
    $K_{AGT}$ & 0.959 & 0.949 & 0.884 & 0.962 & 0.894 & 0.926 & 1.006 & 0.939 & 0.897 & 0.794 & 0.908 & 0.792 & 0.926 & 1.000 \\
    $P_{A}$ & 0.992 & 0.982 & 0.965 & 0.997 & 0.961 & 0.964 & 1.040 & 1.030 & 0.949 & 0.882 & 0.966 & 0.864 & 0.964 & 1.029 \\
    $P_{AGT}$ & 0.965 & 0.960 & 0.925 & 0.972 & 0.924 & 0.941 & 1.017 & 0.969 & 0.922 & 0.847 & 0.933 & 0.830 & 0.941 & 1.012 \\
    \midrule
    $\mathbf{W=1250}$ &       &       &       &       &       &       &       &       &       &       &       &       &       &  \\
    $HAR$ & \textit{0.999} & \textit{0.517} & \textit{0.346} & \textit{0.945} & \textit{0.513} & \textit{0.136} & \textit{0.631} & \textit{1.503} & \textit{0.568} & \textit{0.371} & \textit{1.037} & \textit{0.661} & \textit{0.136} & \textit{0.609} \B \\
    $NN$  & 0.855 & 0.906 & 0.862 & 0.918 & 0.857 & 0.883 & 0.992 & 1.059 & 0.906 & 0.756 & 0.915 & 0.773 & 0.883 & 0.990 \\
    $E_{A}$ & 0.922 & 0.957 & 0.909 & 0.973 & 0.907 & 0.910 & 1.051 & 1.024 & 0.882 & 0.713 & 0.892 & 0.720 & 0.910 & 1.026 \\
    $E_{AGT}$ & 0.929 & 0.939 & 0.860 & 0.955 & 0.866 & 0.903 & 1.020 & 0.958 & 0.866 & 0.672 & 0.881 & 0.692 & 0.903 & 1.005 \\
    $K_{A}$ & 0.936 & 0.952 & 0.890 & 0.969 & 0.892 & 0.921 & 1.030 & 1.061 & 0.887 & 0.710 & 0.902 & 0.715 & 0.921 & 1.009 \\
    $K_{AGT}$ & 0.902 & 0.923 & 0.843 & 0.935 & 0.852 & 0.896 & 1.003 & 0.993 & 0.871 & 0.682 & 0.881 & 0.699 & 0.896 & 0.998 \\
    $P_{A}$ & 0.922 & 0.939 & 0.912 & 0.952 & 0.891 & 0.921 & 1.025 & 1.067 & 0.885 & 0.742 & 0.896 & 0.734 & 0.921 & 1.014 \\
    $P_{AGT}$ & 0.909 & 0.927 & 0.884 & 0.941 & 0.871 & 0.903 & 1.019 & 1.016 & 0.875 & 0.721 & 0.887 & 0.719 & 0.903 & 1.008 \\
    \bottomrule
    \end{tabular}%
    }
  \label{tab:FW_ISOS}%
  \caption{In-sample and out-of-sample results under the fixed-window estimation approach. \enquote{HAR}-labeled rows report actual estimates, all the others, ratios between CV-HAR estimates and their corresponding HAR estimates (the inverse ratio is taken for R2). Values smaller than one indicate that the CV-HAR model outperforms. The first letter in the row names identifies the marginal distribution construction method: ECDF (\enquote{E}), kernel (\enquote{K}), or parametric (\enquote{P}). Subscripts refer to the copulas involved in Vine construction: \enquote{A}-indexed rows refer to Vines allowing for Archimedean copulas only, \enquote{AGT}-indexes indicate that Archimedean, Gaussian, and t- copulas are allowed. Rows indexed with \enquote{NN} refer to ratios wrt. the neural network model.}
\end{table}%

\subsection{Out-of-sample analysis}
\subsubsection{Fixed window}
With respect to the fixed window approach, out-of-sample results from Tab.\ref{tab:FW_ISOS} confirm a general improvement of the ratios favoring the CV-HAR model by the widening of the window size.
Also, the set of pair-copulas, allowing for Archimedeans, Gaussian and t-copula, improves the forecasts over the set allowing for of Archimedean copulas only. 
Similarly, as for in-sample analyses, all the measures at longer windows are supporting the CV-HAR model. There are however exceptions wrt. MSE and R2 measures. In this regard, the best performing window is that of length 750 days. Indeed, as Fig.\ref{fig:2Series} illustrates, volatility at the very beginning of 2017 (the sixth year, i.e. sample days after 1250) is particularly quiet, as never in the preceding sample. Under the longest window, the model closely adapts to the dependence structure earlier observed, becoming gradually rigid in capturing deviations from the actual time series's behavior on which it was estimated. Furthermore, margins may provide an unsatisfactory fit, since such low volatility values have been rarely encountered and the left tail of the modeled distribution can potentially provide a poor fit for the actual data.

The grater rations we uniformly observe for the MSE wrt. to MAE and MAD measures are interpretable by the presence of outliers heavily penalizing the squared loss in the MSE measure wrt. to whose the HAR model appears to be less sensitive. Moreover, as for low quantiles, for rare and very high volatilities, both kernel and parametric distributions could provide an inadequate fitting, since they are extrapolated from a sample that is poorly representative of the actual distribution in the very upper quantiles. However, this discrepancy is mild, based on the detected ratios close to unity.

\subsubsection{Increasing window}
Tab.\ref{tab:OS} reports one-step-ahead forecasts measures for increasing window estimations of the HAR and CV-HAR models. One-step-ahead forecasts are tested to be statistically different by means of the CPA tests. Test statistics and P-values are reported in Tab.\ref{tab:DM_test}.
On a general level, results still favor the CV-HAR model. 
For MAE, MAD, MASE, MAPE and Qlik performance measures, CV-HAR forecasts seem to greatly improve over the HAR ones. 
In general, AGT copulas are preferred, but easier constructions based on Archimedean copulas only are outperforming the HAR model too. It is not the complexity of the copulas involved in the Vine model that drives the performance, but the flexible CV-HAR model itself seems to provide a more attractive alternative for one-step-ahead forecasting wrt. to the linear specification of the HAR model.   

Although MSE ratios are around the unity, by looking at the significance levels of the CPA test (Tab. \ref{tab:DM_test}), differences in MSE between the two models are largely not statistically significant, however unbalanced in favor of the HAR model. As observed for the fixed window analyses, it appears that the squared loss penalizes the CV-HAR specification more than the HAR one. This is not surprising by observing that on low volatility periods the CV-HAR model provides a very satisfactory fit (Fig.\ref{fig:2Series}), while it deteriorates at high volatility regimes. The squared loss shrinks very small residuals and amplifies the large one observed on days of high volatility, leading to an overall squared loss favoring the HAR model. However, the close tracking of the CV-HAR model to the actual volatility observed at low-to-medium volatility days (the vast majority) is much more satisfactory for the CV-HAR model, confirmed both by MAD, MAE, Qlik measures and tests statistics. These are jointly interpretable as an overall ability of the CV-HAR model in forecasting volatility better than HAR in absolute terms, while under a squared loss, high residuals at seldom high-volatility days are covering CV-HAR's overall very satisfactory performance.  Indeed the log-term in the Qlik loss highlights the good performance over the majority of days, as opposed to wide residuals on high-volatility days. Hence, the importance of adopting a wide number of performance measures to unveil such behaviors.
As for the fixed window analyses, the mean directional accuracy (MDA) is similar between the two models: both the models capture the direction of the realized measure with an accuracy of about 62\%. This comment holds for all the sizes of the minimum window. 

\subsubsection{Rolling window}
Lastly, Tab.\ref{tab:OS} reports the results for the rolling window one-step-ahead forecasts, while Tab.\ref{tab:DM_test} reports the DM and CPA test statistics for a significant difference between HAR and CV-HAR forecast errors. 
Tab.\ref{tab:OS} shows a clear cut-off between $W=250$ and wider windows, confirmed by the respective DM statistics. Under this window size, a rolling window approach seems to be not feasible for outperforming the HAR forecasts. This is due to the short data span over which the joint distribution of the volatility terms is modeled via Vine copulas, which, in small-samples, might likely provide an unsatisfactory representation, especially in conveying a proper description of the margins at their tails.
Results look very different for $W=500$ and $W=750$. MAE and Qlik forecasts are statistically different and favoring the HAR model, whereas differences in MSE appear to be non-significant. Other measures such as MAD, MASE, and MAPE clearly support the CV-HAR model as a preferable choice for one-step-ahead volatility forecasting, as for the increasing window approach. The discussion about close-to-unity MSE ratios opposed to below-the-unity MAE and MAD losses for the increasing window estimation applies here as well. On a general level, we observe coherence between the patterns of strong statistical significance for the DM and CPA tests, further validating our analyses.
With AGT pair-copulas there are no major boosts in forecasting performance, suggesting that the simpler modeling estimation involving Archimedean copulas only is a feasible option. Also, models for marginal distributions appear not to play a central role, since holding on a general level: beyond the specific implementation, the CV-HAR alternative seems to capture some non-linear dynamics.

Under RW and IW, the effects of the non-linear flexibility the CV-HAR model appear well-visible. This suggests, along with the in-sample analyses, that the linear specification of the HAR model, through which the conditional expectation of today's volatility given the past terms is implicitly provided by a linear function of the regressors, might be too rigid.

\begin{table}[ht!]
  \centering
  \scalebox{0.75}{
    \begin{tabular}{lccccccc|ccccccc}
          & \multicolumn{7}{c}{Increasing window}                 & \multicolumn{7}{c}{Rolling window} \\
          & MSE   & MAE   & MAD   & MASE  & MAPE  & QLIK  & MDA   & MSE   & MAE   & MAD   & MASE  & MAPE  & QLIK  & MDA \\
    \midrule
    $\mathbf{W=250}$ &       &       &       &       &       &       &       &       &       &       &       &       &       &  \\
    $HAR$ & \textit{1.098} & \textit{0.496} & \textit{0.312} & \textit{0.920} & \textit{0.498} & \textit{0.138} & \textit{0.630} & \textit{1.018} & \textit{0.479} & \textit{0.288} & \textit{0.893} & \textit{0.464} & \textit{0.131} & \textit{0.642} \B \\
    $E_{A}$ & 1.016 & 0.984 & 0.951 & 0.988 & 0.942 & 0.949 & 1.034 & 1.126 & 1.035 & 1.053 & 1.036 & 1.048 & 1.006 & 1.067 \\
    $E_{AGT}$ & 0.989 & 0.962 & 0.889 & 0.968 & 0.899 & 0.933 & 1.015 & 1.076 & 1.011 & 1.003 & 1.012 & 1.008 & 1.000 & 1.044 \\
    $K_{A}$ & 1.040 & 0.990 & 0.968 & 0.994 & 0.941 & 0.983 & 1.031 & 1.084 & 1.023 & 1.050 & 1.025 & 1.034 & 1.016 & 1.055 \\
    $K_{AGT}$ & 0.987 & 0.960 & 0.890 & 0.964 & 0.899 & 0.941 & 1.009 & 1.057 & 1.013 & 1.028 & 1.014 & 1.023 & 1.007 & 1.043 \\
    $P_{A}$ & 1.050 & 0.993 & 0.988 & 0.997 & 0.961 & 0.980 & 1.031 & 1.092 & 1.033 & 1.083 & 1.036 & 1.058 & 1.036 & 1.055 \\
    $P_{AGT}$ & 1.011 & 0.977 & 0.957 & 0.981 & 0.945 & 0.967 & 1.017 & 1.050 & 1.018 & 1.063 & 1.018 & 1.047 & 1.018 & 1.048 \\
    \midrule
    $\mathbf{W=500}$ &       &       &       &       &       &       &       &       &       &       &       &       &       &  \\
    $HAR$ & \textit{1.193} & \textit{0.511} & \textit{0.316} & \textit{0.939} & \textit{0.508} & \textit{0.142} & \textit{0.621} & \textit{1.142} & \textit{0.504} & \textit{0.309} & \textit{0.926} & \textit{0.496} & \textit{0.139} & \textit{0.624} \B \\
    $E_{A}$ & 1.018 & 0.974 & 0.919 & 0.977 & 0.918 & 0.934 & 1.031 & 1.060 & 0.979 & 0.922 & 0.980 & 0.919 & 0.940 & 1.043 \\
    $E_{AGT}$ & 0.989 & 0.961 & 0.874 & 0.966 & 0.891 & 0.919 & 1.018 & 1.034 & 0.965 & 0.875 & 0.966 & 0.890 & 0.935 & 1.025 \\
    $K_{A}$ & 1.045 & 0.982 & 0.942 & 0.985 & 0.918 & 0.977 & 1.028 & 1.084 & 0.980 & 0.931 & 0.983 & 0.916 & 0.960 & 1.031 \\
    $K_{AGT}$ & 0.987 & 0.958 & 0.876 & 0.961 & 0.889 & 0.930 & 1.009 & 1.040 & 0.969 & 0.900 & 0.970 & 0.904 & 0.940 & 1.015 \\
    $P_{A}$ & 1.055 & 0.980 & 0.955 & 0.984 & 0.931 & 0.971 & 1.025 & 1.063 & 0.978 & 0.940 & 0.980 & 0.928 & 0.970 & 1.036 \\
    $P_{AGT}$ & 1.012 & 0.970 & 0.936 & 0.973 & 0.927 & 0.957 & 1.013 & 1.019 & 0.971 & 0.939 & 0.974 & 0.933 & 0.958 & 1.018 \\
    \midrule
    $\mathbf{W=750}$ &       &       &       &       &       &       &       &       &       &       &       &       &       &  \\
    $HAR$ & \textit{1.315} & \textit{0.532} & \textit{0.323} & \textit{0.935} & \textit{0.505} & \textit{0.146} & \textit{0.622} & \textit{1.310} & \textit{0.529} & \textit{0.318} & \textit{0.929} & \textit{0.499} & \textit{0.145} & \textit{0.622} \B \\
    $E_{A}$ & 1.013 & 0.969 & 0.907 & 0.970 & 0.901 & 0.924 & 1.032 & 1.006 & 0.976 & 0.915 & 0.976 & 0.899 & 0.925 & 1.049 \\
    $E_{AGT}$ & 0.988 & 0.956 & 0.855 & 0.959 & 0.868 & 0.908 & 1.021 & 0.986 & 0.960 & 0.860 & 0.961 & 0.862 & 0.915 & 1.030 \\
    $K_{A}$ & 1.042 & 0.977 & 0.924 & 0.979 & 0.901 & 0.968 & 1.026 & 1.032 & 0.968 & 0.902 & 0.971 & 0.886 & 0.960 & 1.030 \\
    $K_{AGT}$ & 0.982 & 0.949 & 0.851 & 0.951 & 0.861 & 0.919 & 1.006 & 1.014 & 0.963 & 0.866 & 0.964 & 0.872 & 0.933 & 1.013 \\
    $P_{A}$ & 1.048 & 0.971 & 0.934 & 0.974 & 0.907 & 0.957 & 1.021 & 1.041 & 0.967 & 0.904 & 0.969 & 0.887 & 0.946 & 1.026 \\
    $P_{AGT}$ & 1.006 & 0.961 & 0.911 & 0.964 & 0.902 & 0.947 & 1.011 & 1.035 & 0.968 & 0.911 & 0.968 & 0.891 & 0.950 & 1.016 \\
    \bottomrule
    \end{tabular}%
    }
      \caption{One-step-ahead forecasts under the increasing window and rolling window estimation schemes. \enquote{HAR}-labeled rows report actual estimates, all the others, ratios between CV-HAR estimates and their corresponding HAR estimates. For information about the labeling refer to Tab.\ref{tab:FW_ISOS}.}
  \label{tab:OS}%
\end{table}%

\section{Conclusion} \label{sec:Concl}
This research shows how the use of a C-Vine copula construction to model the joint distribution of volatility components over different scales can be exploited to model and forecast realized  measures. 
Motivated by the earlier literature investigating the importance of structural breaks and non-linearities within the HAR setting, this paper proposes a non-linear approach that readily exploits the conditional structure arising from the joint distribution of the volatility components involved in the HAR model of \citep{corsi2009simple}. Following a general regression framework, we do not impose any structure on the conditional expectation (i.e. linearity between the variables) but we recover it from the joint (Vine) distribution.  Importantly, this approach guarantees positivity of realized measures' forecasts, standing out as one of the very few approaches (if perhaps not the only one) naturally suitable for non-logarithmic realized measures modeling and forecasting, overcoming, by construction, positivity issues.

This research is inspired by the work of \citet{sokolinskiy2011forecasting}, but it goes beyond their bivariate framework by fully recalling the HAR modeling spirit. We set apart from \citet{sokolinskiy2011forecasting} wrt. marginal distribution modeling, multivariate copula construction, analytic computation of the forecasts (opposed to simulation) and forecasting framework.

We apply the CV-HAR method on real high-frequency financial data, by extracting realized kernel intraday volatility measures for 10 stocks, over almost seven years.
As a general result, the CV-HAR improves all the performance measures considered, in all the three estimation-forecasting schemes examined. A partial exception, applying in particular to small estimation windows, are MSE point-estimates.
Opposed to different performance measures (such as MAD, MAE, Qlike) the quadratic loss seems to be overlooking the very satisfactory performance (small squared residuals) of our approach for the broad majority of low-to-moderate volatility days, while excessively penalizing large losses on days of high volatility.

The CDF modeling plays a secondary role in the results, as well as the copulas involved in the C-Vine construction. Indeed, the CV-HAR specification seems to outperform the HAR alternative, not because of its flexible marginal modeling and copula construction, but rather because the joint distribution modeling approach itself seems to improve over the HAR specification. 
Our results pinpoint that by relaxing original the linear form of the HAR model by allowing for a more general functional linking the volatility components, considerable in-sample and out-of-sample improvements can be achieved. This suggests that the linear form of the HAR model is restrictive, whereas the conditional expectation of today's volatility - given its past terms directly retrieved from their joint distribution, with no functional assumptions - captures a relationship of more complex nature. Under a fixed window approach we report that a simple neural network (NN) seems not to outperform the CV-HAR specification. This suggests that the information conveyed by the joint distribution plays a central role: the non-linear regressor retrieved from the joint distribution of the lagged volatility terms seems preferable over the complex non-linear regressor retrieved with a NN architecture.

Future research may extend the analyses to different realized measures, include D- and R- Vine specifications, and consider a wider set of copula families. 
The CV-HAR model can considered in value-at-risk applications since naturally leading to asymmetric confidence intervals and quantiles around the expected conditional mean.
Importantly, it would be interesting to widen the current analysis against NN alternatives, and extend it to different models where the non-linear functional is retrieved with machine-learning approaches \citep [e.g.][]{lebaron2018forecasting}. This could further shed light on the role of the rich information on the joint distribution of the lagged volatility terms plays in forecasting. 

\section*{Acknowledgments}
The research leading to this manuscript received funding from the European Union's Horizon 2020 research and innovation program under Marie Sk\l{}odowska-Curie grant agreement No. 675044. The author is grateful to Professor Juho Kanniainen, Assoc. Professors Marcelo C. Medeiros and Bezirgen Veliyev for their helpful comments and insightful discussions.

\clearpage
\bibliography{Bib_file}

\newpage
\appendix
\section{Figures and Tables} \label{app:Figures}

\begin{landscape}
\begin{table}[htbp]
  \centering
      \scalebox{0.78}{
   
    \begin{tabular}{rccc|ccc|ccc}
          & \multicolumn{3}{c}{Increasing window} & \multicolumn{6}{c}{Rolling window} \\
          &       &       & \multicolumn{1}{c}{} &       &       & \multicolumn{1}{c}{} &       &       &  \\
          & \multicolumn{3}{c|}{Conditional predictive ability test} & \multicolumn{3}{c}{Diebold-Mariano test} & \multicolumn{3}{c}{Conditional predictive ability test} \\
          & MSE   & MAE   & Qlik  & MSE   & MAE   & \multicolumn{1}{c}{Qlik} & MSE   & MAE   & Qlik \\
    \midrule
    \multicolumn{1}{l}{$\mathbf{W=250}$} &       &       &       &       &       &       &       &       &  \\
    \multicolumn{1}{l}{$E_{A}$} & -3.370 & 10.296 & 48.791 & -2.872 & -4.523 & -0.726 & -11.123 & -20.884 & -4.370 \\
          & (0.185) & (5.812E-03**) & (2.542E-11***) & (4.080E-03**) & (6.140E-06***) & (0.468) & (3.843E-03**) & (2.917E-05***) & (0.113) \\
    \multicolumn{1}{l}{$E_{AGT}$} & 2.580 & 62.577 & 95.609 & -2.956 & -1.745 & -0.094 & -11.734 & -3.093 & -7.437 \\
          & (0.275) & (2.576E-14***) & (***) & (3.127E-03**) & (0.081) & (0.925) & (2.831E-03**) & (0.213) & (0.024*) \\
    \multicolumn{1}{l}{$K_{A}$} & -9.718 & 4.484 & 15.720 & -2.908 & -3.604 & -1.545 & -9.895 & -12.946 & -4.724 \\
          & (7.759E-03**) & (0.106) & (3.858E-04***) & (3.648E-03**) & (3.143E-04***) & (0.122) & (7.101E-03**) & (1.545E-03**) & (0.094) \\
    \multicolumn{1}{l}{$K_{AGT}$} & 2.648 & 95.374 & 85.530 & -2.300 & -2.175 & -0.735 & -10.325 & -5.103 & -12.824 \\
          & (0.266) & (***) & (***) & (0.021*) & (0.030*) & (0.463) & (5.726E-03**) & (0.078) & (1.641E-03**) \\
    \multicolumn{1}{l}{$P_{A}$} & -9.431 & 4.175 & 11.105 & -3.605 & -5.389 & -3.722 & -14.744 & -31.807 & -27.443 \\
          & (8.954E-03**) & (0.124) & (3.878E-03**) & (3.139E-04***) & (7.187E-08***) & (1.982E-04***) & (6.286E-04***) & (1.239E-07***) & (1.099E-06***) \\
    \multicolumn{1}{l}{$P_{AGT}$} & -2.266 & 39.639 & 39.786 & -2.440 & -3.462 & -2.035 & -7.250 & -12.873 & -33.721 \\
          & (0.322) & (2.469E-09***) & (2.293E-09***) & (0.015*) & (5.373E-04***) & (0.042*) & (0.027*) & (1.602E-03**) & (4.759E-08***) \\
    \midrule
    \multicolumn{1}{l}{$\mathbf{W=500}$} &       &       &       &       &       &       &       &       &  \\
    \multicolumn{1}{l}{$E_{A}$} & -3.316 & 18.380 & 59.803 & -1.576 & 2.775 & 6.260 & -8.417 & 7.669 & 39.380 \\
          & (0.191) & (1.020E-04***) & (1.033E-13***) & (0.115) & (5.534E-03**) & (3.990E-10***) & (0.015*) & (0.022*) & (2.811E-09***) \\
    \multicolumn{1}{l}{$E_{AGT}$} & 2.519 & 49.103 & 107.691 & -1.472 & 5.208 & 6.811 & -8.775 & 26.795 & 46.677 \\
          & (0.284) & (2.175E-11***) & (***) & (0.141) & (1.940E-07***) & (1.020E-11***) & (0.012*) & (1.519E-06***) & (7.316E-11***) \\
    \multicolumn{1}{l}{$K_{A}$} & -9.732 & 10.053 & 17.139 & -2.223 & 2.601 & 4.029 & -5.018 & 9.282 & 16.767 \\
          & (7.702E-03**) & (6.561E-03**) & (1.898E-04***) & (0.026*) & (9.300E-03**) & (5.630E-05***) & (0.081) & (9.648E-03**) & (2.287E-04***) \\
    \multicolumn{1}{l}{$K_{AGT}$} & 2.503 & 79.804 & 96.202 & -1.658 & 4.998 & 6.838 & -9.123 & 25.555 & 47.180 \\
          & (0.286) & (***) & (***) & (0.097) & (5.869E-07***) & (8.431E-12***) & (0.010*) & (2.823E-06***) & (5.689E-11***) \\
    \multicolumn{1}{l}{$P_{A}$} & -9.132 & 15.313 & 13.819 & -2.183 & 3.161 & 2.930 & -7.882 & 10.280 & 9.964 \\
          & (0.010*) & (4.729E-04***) & (9.983E-04***) & (0.029*) & (1.578E-03**) & (3.392E-03**) & (0.019*) & (5.858E-03**) & (6.861E-03**) \\
    \multicolumn{1}{l}{$P_{AGT}$} & -2.300 & 49.923 & 47.378 & -1.100 & 5.253 & 5.033 & -5.394 & 28.734 & 27.511 \\
          & (0.317) & (1.443E-11***) & (5.153E-11***) & (0.271) & (1.524E-07***) & (4.912E-07***) & (0.067) & (5.762E-07***) & (1.062E-06***) \\
    \midrule
    \multicolumn{1}{l}{$\mathbf{W=750}$} &       &       &       &       &       &       &       &       &  \\
    \multicolumn{1}{l}{$E_{A}$} & -2.419 & 19.159 & 57.922 & -0.323 & 3.343 & 7.227 & -2.480 & 13.770 & 52.331 \\
          & (0.298) & (6.914E-05***) & (2.645E-13***) & (0.747) & (8.312E-04***) & (5.348E-13***) & (0.289) & (1.023E-03**) & (4.331E-12***) \\
    \multicolumn{1}{l}{$E_{AGT}$} & 2.028 & 44.930 & 94.299 & 0.797 & 5.787 & 7.680 & 2.262 & 34.240 & 58.543 \\
          & (0.363) & (1.753E-10***) & (***) & (0.426) & (7.431E-09***) & (1.766E-14***) & (0.323) & (3.672E-08***) & (1.938E-13***) \\
    \multicolumn{1}{l}{$K_{A}$} & -7.281 & 12.007 & 16.558 & -1.386 & 4.606 & 3.260 & -5.519 & 21.319 & 12.773 \\
          & (0.026*) & (2.470E-03**) & (2.537E-04***) & (0.166) & (4.171E-06***) & (1.117E-03**) & (0.063) & (2.347E-05***) & (1.684E-03**) \\
    \multicolumn{1}{l}{$K_{AGT}$} & 3.416 & 85.132 & 88.411 & -0.685 & 5.503 & 6.971 & -3.849 & 30.536 & 48.325 \\
          & (0.181) & (***) & (***) & (0.493) & (3.842E-08***) & (3.372E-12***) & (0.146) & (2.340E-07***) & (3.209E-11***) \\
    \multicolumn{1}{l}{$P_{A}$} & -7.426 & 21.165 & 16.522 & -1.835 & 4.768 & 4.801 & -4.632 & 26.303 & 23.157 \\
          & (0.024*) & (2.535E-05***) & (2.584E-04***) & (0.067) & (1.891E-06***) & (1.608E-06***) & (0.099) & (1.943E-06***) & (9.367E-06***) \\
    \multicolumn{1}{l}{$P_{AGT}$} & -1.385 & 56.552 & 42.494 & -1.307 & 4.642 & 4.720 & -2.633 & 24.098 & 26.134 \\
          & (0.500) & (5.247E-13***) & (5.923E-10***) & (0.191) & (3.501E-06***) & (2.401E-06***) & (0.268) & (5.850E-06***) & (2.114E-06***) \\
    \bottomrule
    \end{tabular}%
    }
    \caption{Diebold-Mariano and conditional predictive ability tests for one-step-ahead forecasts under increasing window and rolling window estimation schemes. Residual of the CV-HAR model are subtracted from HAR's one: positive DM statistics favours the CV-HAR model. For information about the labeling refer to Tab.\ref{tab:FW_ISOS}. (*) Denotes 5\% significance, (**) 1\% significance and (***) 0.1\% significance.}
  \label{tab:DM_test}%
\end{table}%

\end{landscape}

\end{document}